\documentclass[10pt]{IEEEtran}
\usepackage{cite}
\usepackage{graphicx,subfigure}
\usepackage{amsmath,amssymb}
\usepackage{algorithm, algorithmic}
\usepackage{subfigure}
\usepackage{psfrag}
\usepackage{paralist}
\usepackage{authblk}
\usepackage{yhmath}
\usepackage{multirow}
\usepackage{balance}

\title{Architecture and Algorithms for an Airborne Network$^\dagger$
\thanks{$^\dagger$This research is supported in part by a grant from the U.S. Air Force Office of Scientific Research under grant number FA9550-09-1-0120.}}

\author[1]{Arunabha Sen}
\author[1]{Pavel Ghosh}
\author[1]{Tiffany Silva}
\author[1]{Nibedita Das}
\author[2]{Anjan Kundu}
\affil[1]{Department of Computer Science and Engineering, Arizona State University, Tempe, AZ, 85281 \authorcr \{Pavel.Ghosh, asen, tsilva, nmaulik\}@asu.edu}
\affil[2]{Saha Institute of Nuclear Physics, Kolkata 700064, India \authorcr anjan.kundu@saha.ac.in}

\begin{document}

\maketitle
\begin{abstract}
The U.S. Air Force currently is in the process of developing an Airborne Network (AN) to provide support to its combat aircrafts on a mission. The reliability needed for continuous operation of an AN is difficult to achieve through completely infrastructure-less mobile ad hoc networks. In this paper we first propose an architecture for an AN where airborne networking platforms (ANPs - aircrafts, UAVs and satellites) form the backbone of the AN. In this architecture, the ANPs can be viewed as mobile base stations and the combat aircrafts on a mission as mobile clients. Availability of sufficient control over the movement pattern of the ANPs, enables the designer to develop a topologically stable backbone network. The combat aircrafts on a mission move through a space called {\em air corridor}. The goal of the AN design is to form a backbone network with the ANPs with two properties: (i) the backbone network remains {\em connected at all times}, even though the topology of the network changes with the movement of the ANPs, and (ii) the entire three dimensional space of the air corridor is under {\em radio coverage at all times} by the continuously moving ANPs.

In addition to proposing an architecture for an AN,  the contributions of the paper include, (i) development of an algorithm that finds the velocity and transmission range of the ANPs so that the dynamically changing backbone network remains connected at all times, (ii) development of a routing algorithm that ensures a connection between the source-destination  node pair with the fewest number of path switching, (iii) given the dimensions of the air corridor and the radius of the {\em coverage sphere} associated with an ANP, development of an algorithm that finds the fewest number of ANPs required to provide complete coverage of the air corridor at all times, (iv) development of an algorithm that provides connected-coverage to the air corridor at all times, and (v) results of experimental evaluations of our algorithms, (vi) development of a  visualization tool that depicts the movement patterns of the ANPs and the resulting dynamic graph and the coverage volume of the backbone network.
\end{abstract}

\section{Introduction}
\label{sec:intro}
\begin{center}
\begin{figure*}[tbh]
    \begin{minipage}[tbh]{0.6\linewidth}
        \centering
       	\includegraphics[width=\textwidth, keepaspectratio]{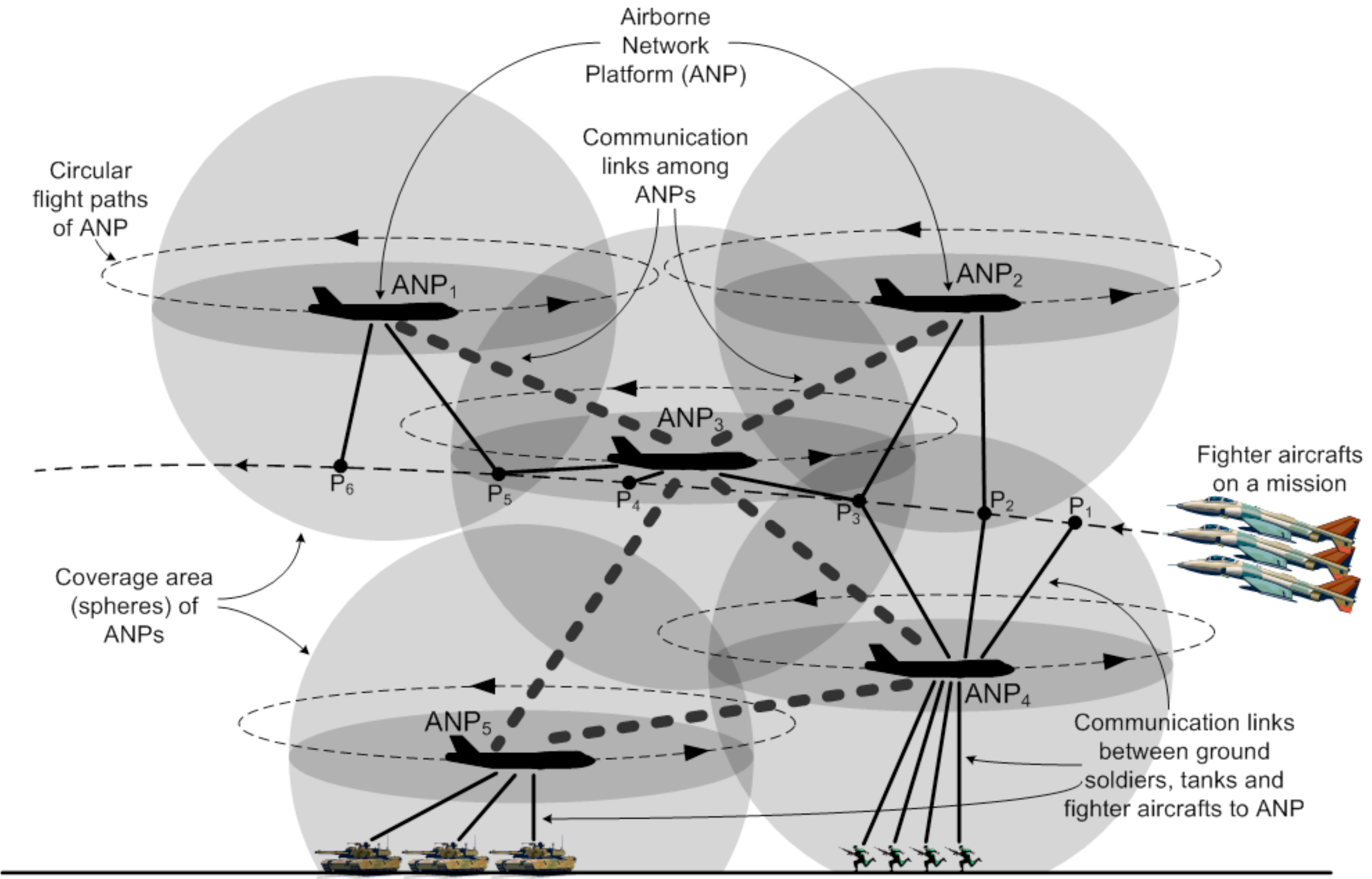}
	\caption{A schematic view of the Airborne Network }
	\label{fig:airborne}
    \end{minipage}
    \hfill
    \begin{minipage}[tbh]{0.35\linewidth}
        \centering
       	\subfigure[Air Corridor and the combat aircrafts on a mission with planned flight paths]{\includegraphics[width=\textwidth, keepaspectratio]{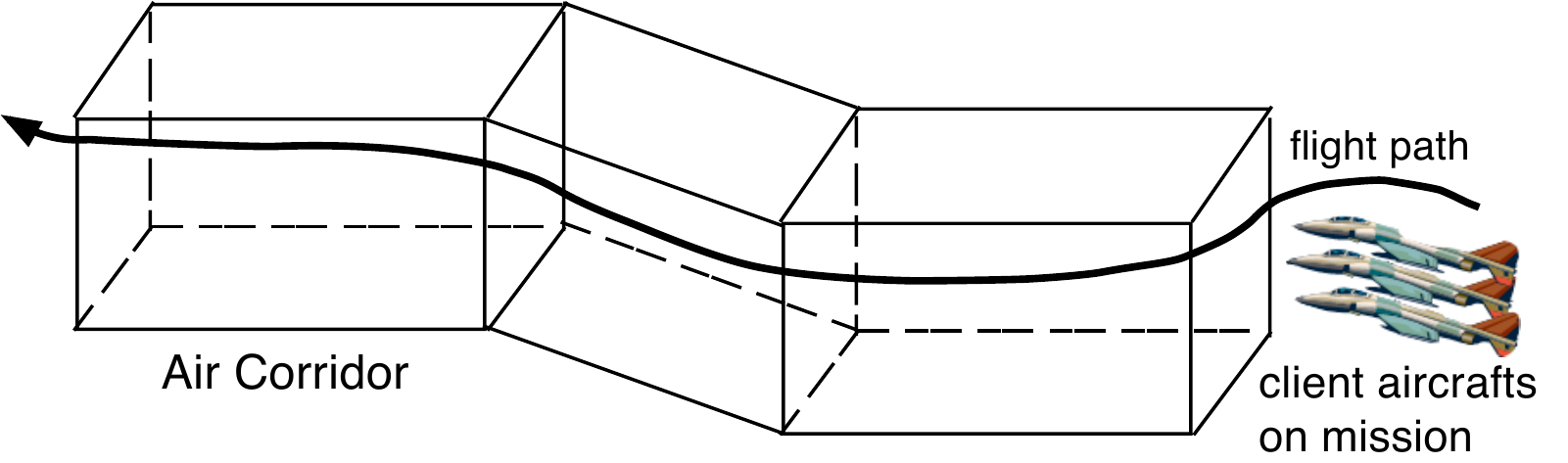}\label{fig:airCorridor1}}
	\hfill
	\subfigure[A section of air corridor]{\includegraphics[width=0.5\textwidth, keepaspectratio]{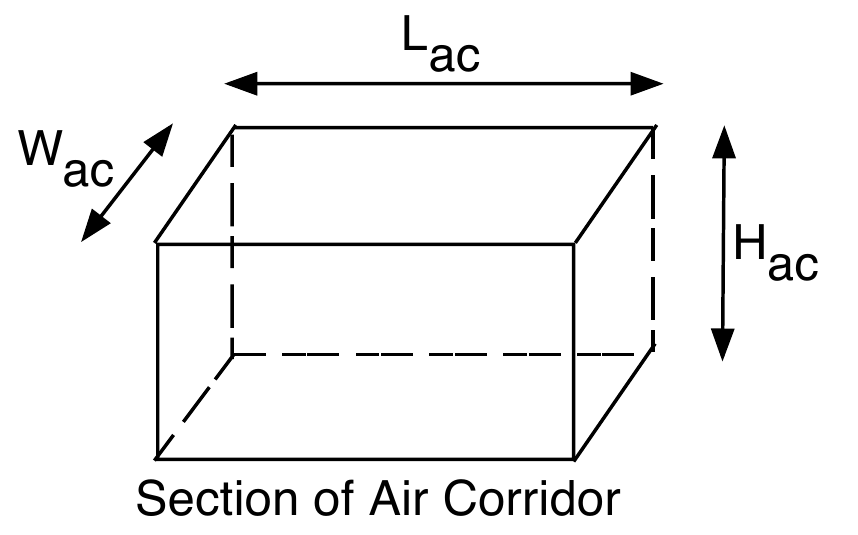}
	\label{fig:airCorridor2}}
	\hfill
	\subfigure[Circular orbits of ANPs (black dots) placed at the top surface of air corridor]{\includegraphics[width=0.4\textwidth, keepaspectratio]{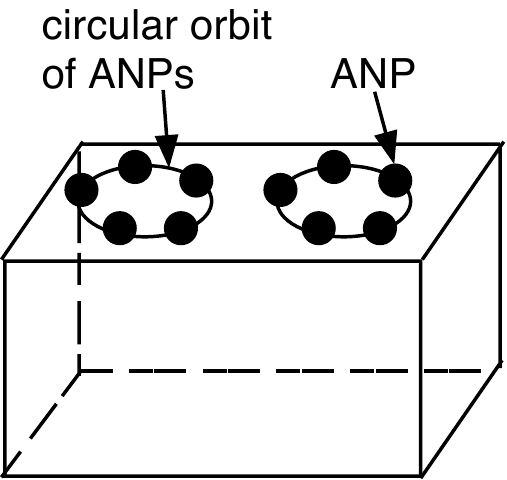}
	\label{fig:airCorridor4}}
	\caption{Air Corridor, rectangular parallelopiped section, client airplanes, ANPs in circular orbits}
	\label{fig:airCorridor}
    \end{minipage}
\end{figure*}
\end{center}

Efforts are currently underway in the U.S. Air Force to utilize a heterogeneous set of physical links (RF, Optical/Laser and SATCOM) to interconnect a set of terrestrial, space and highly mobile airborne platforms (satellites, aircrafts and Unmanned Aerial Vehicles (ANPs)) to form an Airborne Network (AN).
The design, development, deployment and management of a network where the nodes are mobile are considerably more complex and challenging than a network of static nodes. This is evident by the elusive promise of the Mobile Ad-Hoc Network (MANET) technology where despite intense research activity over the last fifteen years, mature solutions are yet to emerge \cite{Burbank06,Con07}. One major challenge in the MANET environment is the unpredictable movement pattern of the mobile nodes and its impact on the network structure. In case of an Airborne Network (AN), there exists considerable control over the movement pattern of the mobile platforms. A senior Air Force official can specify the controlling parameters, such as the {\em location, flight path and speed} of the ANPs to realize an AN with desired functionalities. Such control provides the designer with an opportunity of develop a topologically stable network, even when the nodes of the network are highly mobile. We view the AN as an infrastructure (a wireless mesh network) in the sky formed by mobile platforms such as aircrafts, satellites and UAVs to provide communication support to its clients such as combat aircrafts on a mission. Just as an Airborne Warning and Control System (AWACS) aircraft plays a role in a mission by providing communication support to fighter aircrafts directly engaged in combat, we believe that the aircrafts and ANPs forming the AN will provide similar support to the combat aircrafts over a much larger area. As shown in Fig.~\ref{fig:airCorridor1}, the combat air crafts on a mission fly through a zone referred to as an {\em air corridor}. In addition to forming a connected backbone network, the ANPs are also required to provide complete {\em radio coverage} in the air corridor so that the combat aircrafts, irrespective of their locations within the air corridor, have access to at least one backbone node (i.e., an ANP) and through it, the entire network. Accordingly, the AN is required to have two distinct properties: (1) the backbone network formed by the ANPs must remain {\em connected at all times}, even though the topology of the network changes with the movement of the ANPs, and (2) the entire three dimensional space of the air corridor is {\em covered at all times} by the continuously moving ANPs. To the best of our knowledge this is the first paper that proposes an architecture for an AN and provide solutions for the {\em all time connected-coverage problem of a three-dimensional space with mobile nodes}.

One of the pioneering results in three dimensional coverage problem for sensor networks was presented by Haas {\em et al.} \cite{HAAS08,HAAS06} in which they concluded that the truncated octahedron has the highest volumetric quotient (the ratio of the volume of a polyhedron to the volume of its circumsphere) among all the space filling polyhedrons and utilized this to develop placement strategies for three dimensional underwater sensor networks. Their scheme is a centralized one. Distributed protocol of achieving three dimensional space coverage is found in the research of Tezcan {\em et al.} \cite{TEZ04}. Poduri {\em et al.} \cite{POD06} later on introduced the notion of $NET$ graphs and used it to obtain three dimensional sensor coverage. Similar research aiming at the coverage problem in 3D was also presented by \cite{CHEN08, LEI07, HUANG04}. However none of these researchers put any emphasis on the problem of obtaining coverage while the constituting nodes are mobile in a three dimensional space. The mobile nature of the ANPs in airborne networks add yet another dimension of difficulty to the 3D coverage problem.

In this paper we first propose an architecture for an AN where airborne networking platforms (ANPs - aircrafts, UAVs and satellites) form the backbone or mobile base stations of the AN, and the combat aircrafts on a mission function as mobile clients. We then proceed to determine the the number and initial location of the ANPs, their velocity and transmission range, so that the dynamically changing network retains properties (1) and (2) mentioned in the previous paragraph. The rest of the paper is organized as follows. In Section \ref{sec:sysModel}, we provide the system model and an architecture of an Airborne Network. Section \ref{sec:probFormulation} formally states the connectivity problem for an AN. In Section \ref{sec:connectivity}, we provide an algorithm that finds the velocity and the transmission range of the ANPs so that the dynamically changing  network remains connected at all times. Section \ref{sec:routing} presents a routing algorithm that ensures a connection between the source-destination  node pair with the fewest number of path switching.  Given the dimensions of the air corridor and the radius of the coverage sphere associated with an ANP, Section \ref{sec:coverageProbFormulation} formulates the coverage problem for the air corridor. Section \ref{sec:coverageSolution} presents an algorithm that finds the fewest number of ANPs required to provide complete coverage of the air corridor at all times. The Section \ref{sec:connCoverage} combines results of Sections \ref{sec:connectivity} and \ref{sec:coverageSolution} and presents an algorithm to provide connected-coverage to the air corridor at all times. In Section \ref{sec:visualization} we briefly describe a visualization tool that we developed to demonstrate the movement patterns of the ANPs and its impact on the resulting dynamic graph and the coverage volume of the backbone network. The results of experimental evaluations of our algorithms and related discussion is presented in Section \ref{sec:experiments}. Section \ref{sec:conclusion} concludes the paper.

\section{System Model and Architecture}
\label{sec:sysModel}
A schematic diagram of our view of an AN is shown in Fig.~\ref{fig:airborne}. In the diagram, the black aircrafts are the Airborne Network Platforms (ANP), the aircrafts that form the infrastructure of the AN (although in Fig.~\ref{fig:airborne}, only aircrafts are shown as ANPs,  the UAVs and satellites can also be considered as ANPs). We assume that the ANPs follow a circular flight path. The circular flight paths of the ANPs and their coverage area (shaded spheres with ANPs at the center) are also shown in Fig.~ \ref{fig:airborne}. Thick dashed lines indicate the communication links between the ANPs.  The figure also shows three fighter aircrafts on a mission passing through space known as {\em air corridor}, where network coverage is provided by ANPs 1 through 5. When the fighter aircrafts are at point P1 on their flight path, they are connected to ANP4 because point P1 is covered by ANP4 only. As the fighter aircrafts move along their flight trajectories, they pass through the coverage area of multiple ANPs and there is a smooth hand-off from one ANP to another when the fighter aircrafts leave the coverage area of one ANP and enter the coverage area of another. The fighter aircrafts are connected to an ANP as long as they are within the coverage area of that ANP. At points P1, P2, P3, P4, P5 and P6 on their flight path in Fig.~\ref{fig:airborne}, the fighter aircrafts are connected to the ANPs (4), (2, 4), (2, 3, 4), (3), (1, 3) and (1), respectively.

One major difference between the wireless mesh networks deployed in many U.S. cities \cite{Wifi08} and the ANs is the fact that, while the nodes of the wireless mesh networks deployed in the U.S.
cities are static, the nodes of an AN are highly mobile. However, as noted earlier, the AN designer has considerable control over the movements of the mobile platforms forming the AN. She can decide on the {\em locality} where the aircraft/ANPs should fly, its {\em altitude, flight path} and {\em speed of movement}. Control over these four important parameters, together with the knowledge of the {\em transmission range} of the transceivers on the flying platforms, provides the designer with an opportunity for creating a fairly stable network, even with highly mobile nodes. In this paper, we make a simplifying assumptions that two ANPs can communicate with each other whenever the distance between them does not exceed the specified threshold (transmission range of the onboard transmitter).  We are well aware of the fact that successful communication between two airborne platforms depends not only on the distance between them, but also on various other factors such as (i) the line of sight between the platforms \cite{TIW08}, (ii) changes in the atmospheric channel conditions due to turbulence, clouds and scattering, (iii) the  banking angle,  the wing obstruction and the dead zone  produced by the wake vortex of the aircraft \cite{EPS04} and (iv) Doppler effect \cite{Doppler} . Moreover, the transmission range of a link is not a constant and is impacted by various factors, such as transmission power, receiver sensitivity, scattering loss over altitude and range, path loss over propagation range, loss due to turbulence and the transmission aperture size \cite{EPS04}. However, the distance between the ANPS remains a very important parameter in determining whether communication between the ANPs can take place, and as the goal of this research is to understand the basic and fundamental issues of designing an AN with twin invariant properties of coverage and connectivity, we feel such simplifying assumptions are necessary and justified. Once the fundamental issues of the problem are well understood, factors (i) through (iv) can be incorporated into the model to obtain a more accurate solution.

\section{Design for Connectivity - Problem Formulation}
\label{sec:probFormulation}
It is conceivable that even if the network topology changes due to movement of the nodes, some underlying structural properties of the network may still remain invariant.
A structural property of prime interest in this context is the {\em connectivity} of the dynamic graph formed by the ANPs. We want the ANPs to fly in such a way, that even though the links between them are established and disestablished over time, the underlying graph remains connected at all times. Although we give connectedness of the graph as an example of a structural property, many other graph theoretic properties $\mathbf{P}$ can be specified as design requirements for the network. The problem can be described formally in the following way.

Consider $n$ nodes (flying platforms) in an $m$-dimensional space ${\mathbb{R}}^m$ (for ANP network scenario $m = 3$). We denote by $x_i(t) \in {\mathbb{R}}^m$ the coordinates of the node $i$ at time $t$, where by convention $x_i$ is considered a $m \times 1$ column vector, and by $\mathbf{x}(t) = {[{x_1}^T(t), \ldots, {x_n}^T(t)]}^T$, the $mn \time 1$ vector resulting from stacking the coordinates of the
nodes in a single vector. Suppose that the dynamics of node $i$ (for all $i \in \{1, 2, \ldots, n\}$), is given by $\dot{x}_i(t) = u_i(t)$, where $u_i(t)$ is the control vector taking values in some set $U \subseteq {\mathbb{R}}^m$. In vector notation, the system dynamics become
\begin{equation}
\mathbf{\dot{x}}(t) = {\mathbf{u}}(t)
\end{equation}

where ${\mathbf{\dot{x}}}(t) = [{\dot{x}_1}^T(t), \ldots, {\dot{x}_n}^T(t)]^T$ and $\mathbf{u}(t) = [{u_1}^T(t), \ldots, {u_n}^T(t)]^T$ are $mn \times 1$ vectors, respectively. The network of flying platforms described by system (1), gives rise to a {\em dynamic graph} ${\cal G}(\mathbf{x}(t))$.

${\cal G}(\mathbf{x}(t)) = ({\cal V}, {\cal E}({\mathbf{x}}(t)))$ is a dynamic graph consisting of

\begin{itemize}
\item a set of nodes ${\cal V} = \{1, 2, \ldots, n\}$ indexed by the set of flying platforms, and\vspace{-0.0in}
\item a set of edges ${\cal E} (\mathbf{x}(t)) = \{(i, j) \mid d_{ij} (\mathbf{x}(t)) < \delta\}$ with $d_{ij}(\mathbf{x}(t)) = \parallel{x_i(t) - x_j(t)}\parallel$ as the Euclidean distance between the platforms $i$ and $j$ and $\delta >0$ is a constant.
\end{itemize}

Since we have control over the node dynamics, the question that naturally arises is whether we can control the motion of the ANPs so that ${\cal G}(\mathbf{x}(t))$ retains graph-theoretic properties of interest $\mathbf{P}$ for all time $t > 0$. A graph $\cal G$ is connected if there exists a path between any two nodes of the graph $\cal G$. Often times the property $\mathbf{P}$ will correspond to the requirement that the graph $\cal G$ remains connected at all times.  Formally the problem can be stated as follows. Suppose that ${\cal C}_{n,\mathbf{P}}$ is the set of all graphs on $n$ nodes with property $\mathbf{P}$. Is it possible to find a control law $\mathbf{u}(t)$ such that if ${\cal G}(\mathbf{x}(0)) \in {\cal C}_{n,\mathbf{P}}$ then ${\cal G}(\mathbf{x}(t)) \in {\cal C}_{n,\mathbf{P}}$ for all $t \geq 0$?

Although a few researchers have studied problems in this domain \cite{Mesbahi051, Mesbahi052, Zav07}, many important questions still remain unanswered. For example, in our study of the movement pattern of the ANPs to create a connected network, we assume that the flight paths of the mobile platforms are already known and we want to find out the speed at which these platforms should move, so that the resulting dynamic graph remains connected at all times. The studies undertaken in \cite{Mesbahi051,Mesbahi052,Zav07} do not address such issues. Although the movement of the airborne platforms will be in a three dimensional space, in a  simplified version of the problem in two dimension (i.e., when all the aircrafts are flying at the same altitude) the problem can be stated as follows:

\noindent
{\em Mobility Pattern for Connected Dynamic Graph  (MPCDG)}: This problem has five controlling parameters:\\
(i) a set of points \{$p_1, p_2, \ldots, p_n\}$ on a two (or three) dimensional space (representing the centers of circular flight paths of the platforms),\\ (ii) a set of radii \{$r_1, r_2, \ldots, r_n$\} representing the radii of circular flight paths, \\(iii) a set of points \{$l_1, l_2, \ldots, l_n$\} representing the initial locations (i.e., locations at time $t = 0$) of the platforms on the circular flight paths, \\ (iv) a set of velocities \{$v_1, v_2, \ldots, v_n$\} representing the speeds of the platforms, and\\ (v) transmission range $T_r$ of the transceivers on the airborne platforms.

\section{Design for Connectivity - Solution}
\label{sec:connectivity}

In the MPCDG problem scenario, any structural property {\bf P} of the resulting dynamic graph will be determined by the problem parameters (i) through (v). The problems that arise in this formulation are as follows:  Given any four of the five problem parameters, how to determine the fifth one, so that the resulting dynamic graph retains property {\bf P} at all times?
Most often we would like to know that given (i), (ii), (iii) and (v), at what speed the ANPs should fly so that the resulting graph is {\em connected} at all times. Alternately, we may want to determine the minimum transmission range of the ANPs to ensure connectivity. In this case, the problem will be specified in the following way. Given (i), (ii), (iii) and (iv), what is the minimum  transmission range of the ANPs so that the resulting graph is {\em connected} at all times? In order to answer these questions, we first need to able to answer a simpler question. Given all five problem parameters including the speed of the ANPs, how do you determine if the resulting dynamic graph is connected at all times? We discuss this problem next.

\subsection {Connectivity checking for AN when all control parameters specified}

\begin{figure}[!t]
 \centering
\includegraphics[width=0.45\textwidth,keepaspectratio]{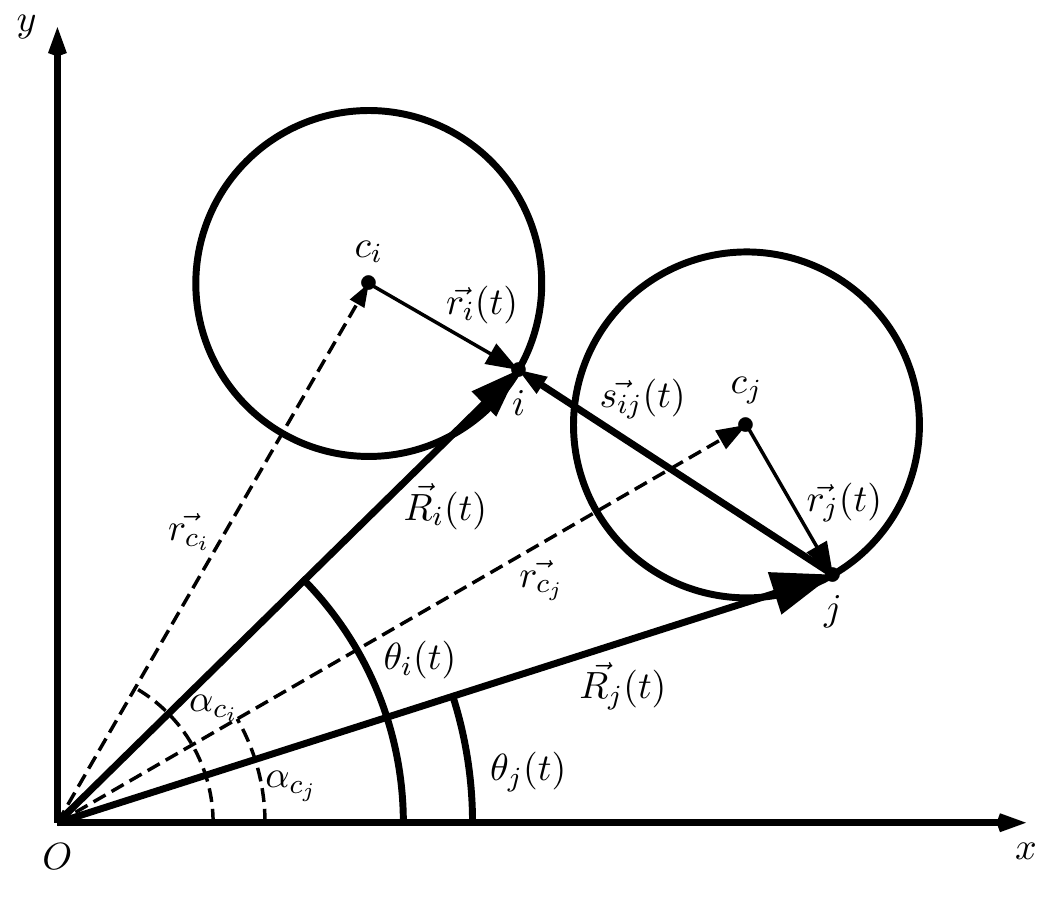}
\caption{Vector representations ($\vec{R_{i}}(t)$ and $\vec{R_{j}}(t)$) of two points $i$ and $j$ at time $t$ moving along two circular orbits: $r_{c_{i}}=15,~r_{c_{j}}=27,~\angle{c_{i}Ox}=\alpha_{c_{i}}=\frac{\pi}{3},~\angle{c_{j}Ox}=\alpha_{c_{j}}=\frac{\pi}{6}$}
\label{fig:pavelpolar}
\end{figure}

In this subsection we describe our technique to find answer to the question posed in the previous paragraph. Suppose that two ANPs, represented by two points $i$ and $j$ (either in two or in three dimensional space, the two dimensional case corresponds to the scenario where the ANPs are flying at same altitute) are moving along two circular orbits with centers at $c_{i}$ and $c_{j}$ with orbit radius $r_{i}$ and $r_{j}$  as shown in Fig.~\ref{fig:pavelpolar} with velocities $v_{i}$ and $v_{j}$ (with corresponding angular velocities $\omega_{i}$ and $\omega_{j}$), respectively.
\begin{figure}[!t]
\centering
\includegraphics[width=0.35\textwidth,keepaspectratio]{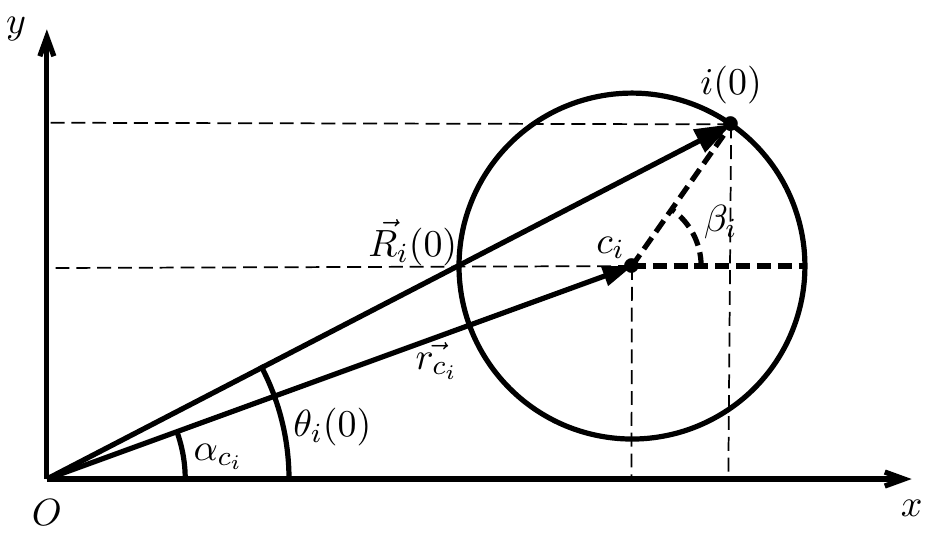}
\caption{Initial phase angle $\beta_{i}$ of point $i$; at time $0$ point is shown as $i(0)$}
\label{fig:betafig}
\end{figure}

A moving  node $i$ is specified by the radius vector $\vec{R_{i}}(t)$ directed from some origin point $O$, and similarly $\vec{R_{j}}(t)$ for point $j$. Therefore the distance $s_{ij}(t)$ between the nodes $i-j$ at  time $t$ is given by:
\begin{equation}
s^{2}_{ij}(t)=(\vec R_{i}(t)-\vec R_{j}(t))^{2}= R_{i}^{2}(t) +  R_{j}^{2}(t) - 2 \vec R_{i}(t) \cdot \vec R_{j}(t)
 \label{eq:vs}
 \end{equation}
As mentioned earlier, we have assumed that the communication between the ANPS is possible if and only if the Euclidean distance between them does not exceed the communication threshold distance $D$. This implies that the link between the nodes $i$ and $j$ is alive (or active) when 
\begin{equation}
s_{ij}(t) \leq D
\label{eq:sD}
\end{equation}

\medskip
\noindent
In the analysis that follows, we have assumed that ANPs are flying at the same altitude, i.e., we focus our attention to the two dimensional scenario. However, this analysis can easily be extended to the three dimensional case to model the scenario where the  ANPs are flying at different altitude. In this case we can view the ANPs as points on a two-dimensional plane moving along two circular orbits, as shown in Fig. \ref{fig:pavelpolar}.  In Fig.~\ref{fig:pavelpolar}, the vectors from the origin $O$ to the centers of the orbits $c_{i}$ and $c_{j}$ are given as $\vec{r_{c_{i}}}$ and $\vec{r_{c_{j}}}$. The cartesian co-ordinates of the centers can be readily obtained as $\vec{r_{c_{i}}} = (r_{c_{i}}cos~\alpha_{c_{i}}, r_{c_{i}}sin~\alpha_{c_{i}})$ and $\vec{r_{c_{j}}} = (r_{c_{j}}cos~\alpha_{c_{j}}, r_{c_{j}}sin~\alpha_{c_{j}})$. Accordingly, $\vec{R_{i}}(t)$ can be expressed in polar coordinates: $R_{i}(t) ,\theta_{i}(t)$ with respect to origin point $O$, as shown in Fig.~\ref{fig:pavelpolar}, and similarly for $\vec{R_{j}}(t)$. The initial location of the points $\vec{R_{i}}(0)$ and $\vec{R_{j}}(0)$ are given. From Fig.~\ref{fig:betafig}, the phase angle $\beta_{i}$ for node $i$ with respect to the center of orbit $c_{i}$, can be calculated as (by taking projection on the axes):
\begin{equation}
tan~\beta_{i} = \frac{ R_{i}(0)cos~\theta_{i}(0) - r_{c_{i}}cos~\alpha_{c_{i}} }{ R_{i}(0)sin~\theta_{i}(0) - r_{c_{i}}sin~\alpha_{c_{i}}}
\label{eq:beta}
\end{equation}

\noindent
Since from Fig.~\ref{fig:pavelpolar},
\begin{equation}
\vec R_{i}(t)=\vec r_{c_i} + \vec r_{i}(t)
\label{eq:vR}
\end{equation}
where $\vec r_{i}(t) = (r_{i} \cos~(\beta_{i} + \omega_{i}t ), r_{i} \sin~(\beta_{i} + \omega_{i}t ))$ (since angle made by $i$ at time $t$ w.r.t. $c_{i}$ is given by $(\beta_{i} + \omega_{i}t)$). Therefore, the angle between $\vec r_{i}(t)$ and $\vec r_{c_{i}}$ is $(\beta_{i} - \alpha_{c_{i}} + \omega_{i}t )$. Hence, 

\begin{equation}
R_{i}^{2}(t)= r_{c_i}^2 + r_{i}^{2} + 2r_{c_i} r_{i} \cos~(\beta_{i} - \alpha_{c_{i}} + \omega_{i}t )
\label{eq:R}
\end{equation}

\noindent
Now taking the projection of  $\vec R_{i}(t)=\vec r_{c_i} + \vec r_{i}(t)$ on the $x$ and $y$ axes, we get
\begin{eqnarray}
R_{i}(t)\cos \theta_{i}(t) & = & r_{c_i}\cos~\alpha_{c_i} + r_{i}\cos~(\beta_{i} + \omega_{i}t),
\label{eq:Rx}
\\
R_{i}(t)\sin~\theta_{i}(t) & = & r_{c_i}\sin~\alpha_{c_{i}} + r_{i}\sin~(\beta_{i} + \omega_{i}t)~~~
\label{eq:Ry}
\end{eqnarray}

\noindent
Recalling $\cos (A-B) = \cos A \cos B + \sin A \sin B$, and simplifying we get
\begin{eqnarray}
&&R_{i}(t)R_{j}(t)\cos (\theta_{i}(t)-\theta_{j}(t)) = r_{c_i} r_{c_j} \cos~\alpha_{c_{i}c_{j}}\nonumber \\
&& + r_{i} r_{j} \cos(\beta_{ij}+(\omega_{i}-\omega_{j})t)+r_{c_i}r_{j} \cos(\alpha_{c_{i}} - \beta_{j} - \omega_{j}t) \nonumber \\
&& + r_{c_j}r_{i} \cos(\alpha_{c_{j}} - \beta_{i} - \omega_{i}t)
\label{eq:cosij}
\end{eqnarray}
where $\alpha_{c_{ij}}=\alpha_{c_{i}}-\alpha_{c_{j}}$ and $\beta_{ij} = \beta_{i} - \beta_{j}$. Combining equation \ref{eq:vs} with equations \ref{eq:R} and \ref{eq:cosij}, we have:

\begin{eqnarray}
s_{ij}^{2}(t) &= & r_{c_{i}}^{2} + r_{i}^{2} + 2r_{c_{i}}r_{i}\cos(\beta_{i} - \alpha_{c_{i}}+ \omega_{i}t) \nonumber\\
& + & r_{c_{j}}^{2} + r_{j}^{2} + 2r_{c_{j}}r_{j}\cos(\beta_{j} - \alpha_{c_{j}} + \omega_{j}t) \nonumber \\
& + & r_{c_i} r_{c_j} \cos~\alpha_{c_{i}c_{j}} + r_{i} r_{j} \cos(\beta_{ij} +(\omega_{i}-\omega_{j})t) \nonumber \\
& + & r_{c_i}r_{j} \cos(\alpha_{c_{i}} - \beta_{j} - \omega_{j}t) \nonumber \\
& + & r_{c_j} r_{i} \cos(\alpha_{c_{j}} - \beta_{i} - \omega_{i}t)
\label{eq:finals1}
\end{eqnarray}

In equation \ref{eq:finals1}, all parameters on the right hand side are known from the initial state of the system, and thus the distance $s_{ij}(t) $ between the nodes  $i-j$ at any time $t$ can be obtained. If the ANPs move at same velocity, i.e., $\omega_{i}=\omega_{j} = \omega$ for all $i,j$ and the radius of the circular orbits are identical, i.e., $r_{i} = r_{j} = r$ for all $i,j$, and the above expression simplifies to:
\begin{eqnarray}
s_{ij}^{2}(t) &= & r_{c_{i}}^{2} + r^{2} + 2r_{c_{i}}r\cos(\beta_{i} - \alpha_{c_{i}}+ \omega t) \nonumber\\
& + & r_{c_{j}}^{2} + r^{2} + 2r_{c_{j}}r\cos(\beta_{j} - \alpha_{c_{j}} + \omega t) \nonumber \\
& + & r_{c_i} r_{c_j} \cos~\alpha_{c_{i}c_{j}} + r^{2} \cos\beta_{ij} \nonumber \\
& + & r_{c_i}r \cos(\alpha_{c_{i}} - \beta_{j} - \omega t) \nonumber \\
& + & r_{c_j}r \cos(\alpha_{c_{j}} - \beta_{i} - \omega t)
\label{eq:finals2}
\end{eqnarray}

\begin{figure*}[htb]
\centering
\subfigure[Distance between two points $i$ and $j$ as a function of time]{
\includegraphics[width=0.48\textwidth,keepaspectratio]{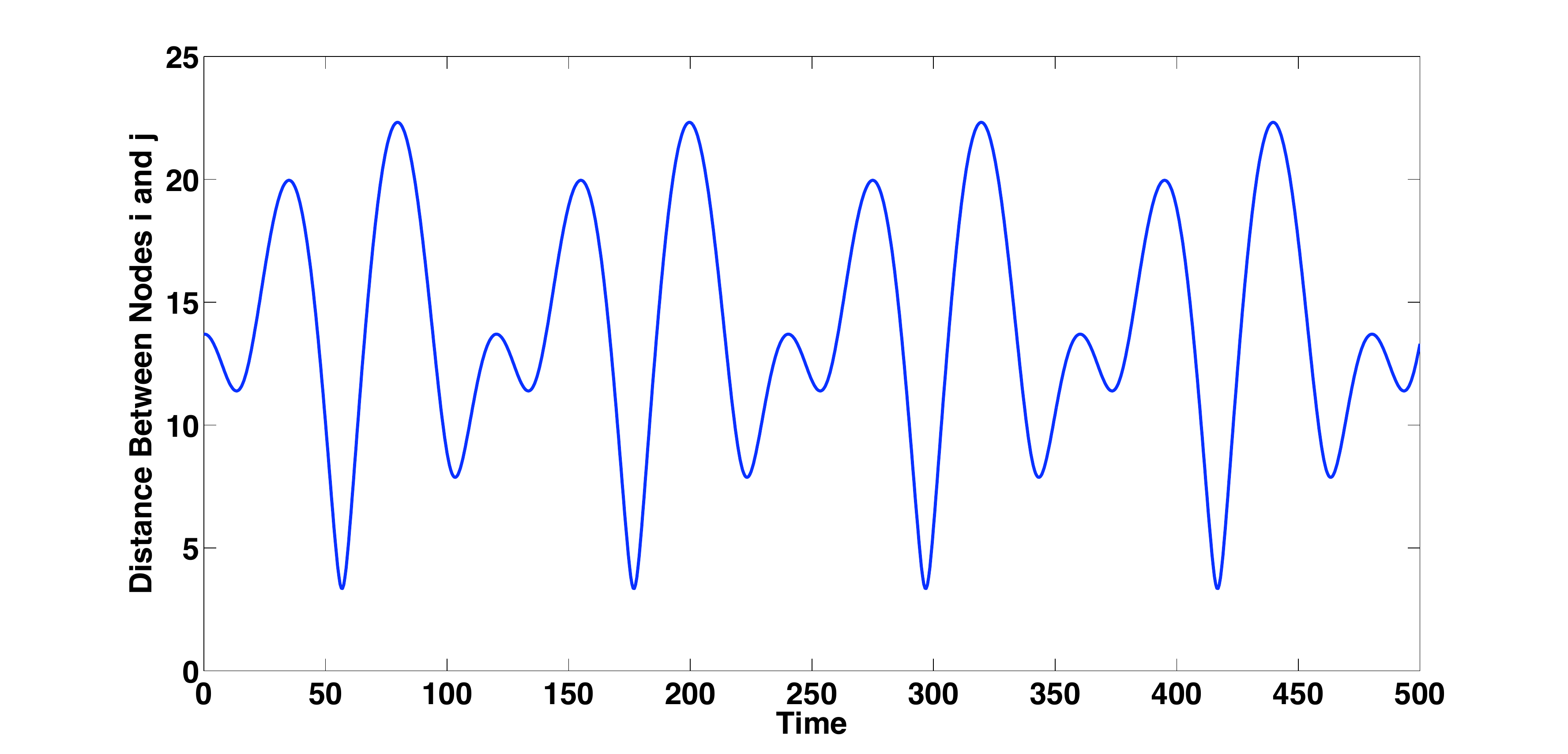}
\label{fig:waveform1}
}
\hfill
\subfigure[Active (Blue)/Inactive (Red) times of the link between $i$ and $j$ with transmission range $T_r$ = 18 ]{
\includegraphics[width=0.48\textwidth,keepaspectratio]{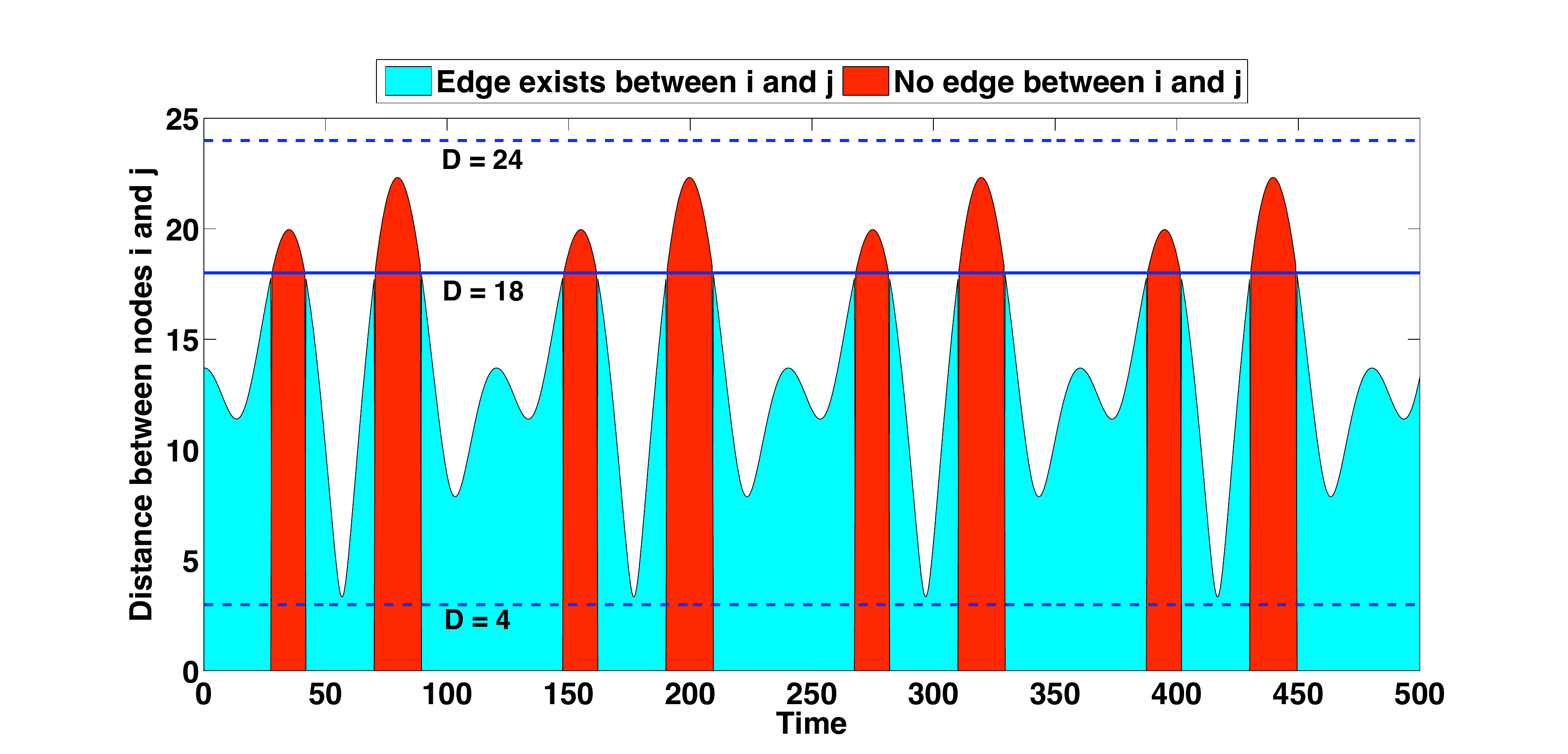}
\label{fig:waveform2}
}
\caption{Effect of the distance between nodes on the existence of the communication link between them}
\label{fig:waveform}
\end{figure*}

If the problem parameters (i) through (v) are specified, we can check if the dynamic graph is connected at all times following these two steps. In the first step, we determine the lifetime (active/inactive) of a link between a pair of nodes $i$ and $j$ in the following way.

\vspace{0.1 in}
\noindent
{\bf Algorithm 1:  Link Lifetime Computation }

\vspace{0.05 in}
\begin{tabbing}
123\=123\=12\=12\=123\=123\=123\= \kill
1. {\em begin}\\
2. \> Using equation (\ref{eq:finals1}), compute and plot the distance\\
\> \> between a pair nodes $i$ and $j$, $s_{ij}(t)$, as a function\\
\> \> of time; (\* See Fig.~\ref{fig:waveform1})\\
3. \> Draw a horizontal line in the $s_{ij}(t)$ versus $t$ plot with\\
 \> \> $s_{ij}(t) = D$, where $D$ is the communication threshold,\\
 \> \> i.e., communication between $i$ and $j$ is possible\\
 \> \>  if $s_{ij}(t) \leq  D$ and impossible otherwise.\\
 \> \> Call this line {\em communication threshold line}, CTL.\\
4. \> The CTL is divided in to segments corresponding\\
\> \>  to the parts where $s_{ij}(t) \leq  D$ and where $s_{ij}(t) >  D$.\\
5. \> Projections of the CTL segments on the $x$-axis \\
\> \> (i.e., the time line) indicates the times when\\
\> \> the link between $i$ and $j$ is alive and \\
\> \>  when it is not. (\* See Fig.~\ref{fig:waveform2})\\
6. {\em end}
\end{tabbing}

Using Algorithm 1, we can compute the life time of every link (i.e., every pair of nodes) in the network. In step 2, using Algorithm 2 (given below) we divide the time line into smaller intervals and determine exactly the links that are active during each of these interval. For each of the intervals we check if the AN graph is connected during that interval using connectivity checking algorithm in \cite{Cormen}. The algorithm is described in detail next.

\vspace{0.1 in}
\noindent
{\bf Algorithm 2: Checking Connectivity of Airborne Network \\
between time $t = t_1$ and $t = t_2$}

\vspace{0.05in}
\begin{tabbing}
123\=123\=12\=12\=123\=123\=123\= \kill
1. {\em begin}\\
2. \> Using the Algorithm for Link Lifetime Computation, \\
\> \> compute the lifetimes of links between all node pairs\\
\> \> and plot them over time line. (See Fig.~\ref{fig:interval})\\
3. \> Draw a vertical line through start and finish\\
\> \>  time of each interval associated with a link on the\\
\> \>  $x$-axis (time line)\\
4. \> Repeat step 3 for each link of the network\\
5. \> The $x$-axis (time line) is divided into a number of \\
\> \> smaller intervals. (See Fig.~\ref{fig:interval}, intervals are \\
\> \> numbered from 1 through 17). From the figure,\\
\> \> we can identify all the links that are alive\\
\> \>  during any one interval. \\
6. \>  Check if the AN graph is connected with the set \\
\> \> of live links during one interval. This can be done\\
\> \> with the connectivity testing algorithm in \cite{Cormen}\\
7. \> Repeat step 6 for all the intervals between $t = t_1$ \\
\> \> and $t = t_2$.\\
8. \> If the AN graph remains connected for all intervals, \\
\> \> conclude that the AN remains connected during\\
\> \>  the entire duration between $t = t_1$ and $t = t_2$, \\
\> \> otherwise conclude that the specified problem\\
\> \> parameters does not ensure a AN that remains \\
\> \> connected during the entire time interval \\
\> \> between  $t = t_1$ and $t = t_2$.\\
6. {\em end}
\end{tabbing}

An example of a plot of equation (\ref{eq:finals1}) (generated using MATLAB) is shown in Fig.~\ref{fig:waveform1} with  communication threshold distance $D =18$. This implies that the link between the nodes $i$ and $j$ exists, when the distance between them is at most $18$ and the link does not exist otherwise. This is shown in Fig.~\ref{fig:waveform2}. The red part indicates the time interval when the link is {\em inactive}(or dead) and the blue part indicates when it is {\em active} (or live). 

Thus using equation (\ref{eq:finals1})  and comparing the distance between any two nodes with the threshold distance $D$, we can determine active/inactive times of all links. This can be represented as intervals on a time line as shown in Fig.~\ref{fig:interval}.  By drawing projections from the end-points of the active/inactive times of each link on the time line, we can find out all the links that are active during a interval on time line. As shown in Fig.~\ref{fig:interval}, links 1, 2 and 3 are active in interval 1; links 1 and 3 are active in interval 2, links 1, 2 and 3 are active in interval 3 and so on. Once we know all the links that are active during a time interval, we can determine if the graph is connected during that interval using any algorithm for computing graph connectivity \cite{Diestel05}. By checking if the graph is connected at all intervals, we can determine if the graph is connected at all times, when the ANPs are moving at specified velocities.

\subsection {Finding the velocity of the ANPs to ensure a connected AN during operational time between $t = t_1$ and $t = t_2$}

In subsection $A$, we have described a technique to determine if the AN remains connected during the entire operational time between $t = t_1$ and $t = t_2$), when all problem parameters, (i)-(v) are specified. In this subsection we try to determine the problem parameter (iv) (i.e., velocity of the ANPs) that we ensure a connected AN during the entire operational time between $t = t_1$ and $t = t_2$) when all other problem parameters have already been specified. The minimum and maximum operating velocities of the ANPs ($v_{min}, v_{max}$) are known. By conducting a binary search on this range, we can compute the minimum velocity at which the ANPs should fly, so that the AN remains connected during the entire operational time. Alternately, we can also try to determine the velocity at which the ANPs should fly, so that the AN remains connected during the entire operational time and {\em fuel consumption by the ANPs is minimized}. If it is known that the fuel consumption is minimized when the ANPs fly with velocity $v_{optimal}$, we can find the velocity that is closest to $v_{optimal}$ and also ensures connectivity of the AN during entire operational time by a targeted search within the range ($v_{min}, v_{max}$). 

\subsection {Finding the transmission range of the ANPs to ensure a connected AN during operational time between $t = t_1$ and $t = t_2$}

In this subsection we try to determine the problem parameter (v) (i.e., transmission range of the ANPs) that we ensure a connected AN during the entire operational time between $t = t_1$ and $t = t_2$) when all other problem parameters have already been specified. The maximum transmission range of an ANP is known in advance ($T_{max}$. By conducting a binary search within the range $0 - T_{max}$, we can determine the the smallest transmission range that will ensure a connected AN during the entire operational time when all other problem parameters have already been determined.

\begin{figure}[!t]
\centering
\includegraphics[width=0.48\textwidth, keepaspectratio]{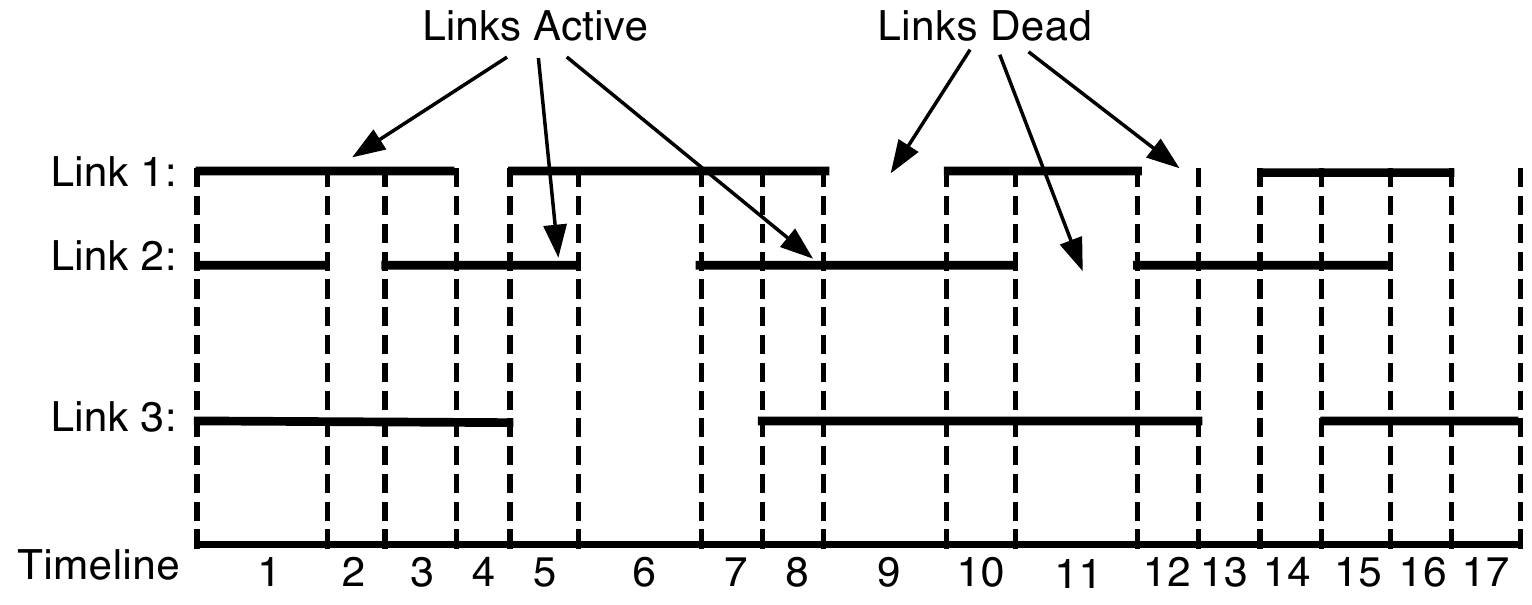}
\caption{Active/Inactive time interval of each link and interval intersection projections on the time line}
\label{fig:interval}
\end{figure}

\section{Routing with Minimum Path Switching}
\label{sec:routing}

In the previous section we described a procedure to determine the velocity of the ANPs so that the resulting dynamic graph is connected at all times. Although the graph remains connected at all times, as the links come and go (alive or dead) a path between a source-destination pair  may not exist for the entire duration of communication. Suppose that a node $s$ has to communicate with another node $d$ from time $t = 5$ to $t = 12$.  Since the graph is connected at all times,  at least a path, say $P_1$, exists from $s$ to $d$ at $t = 5$. However, this path may not exist till $t = 12$. Suppose that as one of its link dies, $P_1$ breaks at $t = 7$. Clearly $P_1$ cannot be used for communication between $s$ and $d$ at $t = 7$. Since the graph is connected at all times, there must exist at least one path, say $P_2$, between $s$ and $d$ at $t = 7$. Therefore data  can be transferred  from $s$ to $d$ using $P_2$ at $t = 7$. However, $P_2$ can break at $t = 10$, in which case a yet another path, say $P_3$ (which is guaranteed to exist because the graph is connected at all times)  can be used for communication between $s$ and $d$ from $t= 10$ to $t = 12$. In such a scenario, the path sequence $P_1 \rightarrow P_2 \rightarrow P_3$ is used for communication between $s$ and $d$ in the time interval $t = 5$ to $t = 12$. In this scenario the path has to be {\em switched} {\em two times}, once from  $P_1 \rightarrow P_2$  and the other time from $P_2 \rightarrow P_3$.  However, it is possible that communication between $s$ and $d$ in the time interval $t = 5$ to $t = 12$ could have been achieved by only {\em one} path switching, using a path $P_4$ from $t = 5$ to $t = 9$ and a path $P_5$ from $t = 9$ to $t = 12$. Since path switching involves a certain amount of overhead, it is undesirable and as such we would like to accomplish routing for the duration of communication with {\em as few path switching as possible}.

\begin{figure}[!t]
\centering
\includegraphics[width=0.45\textwidth,keepaspectratio]{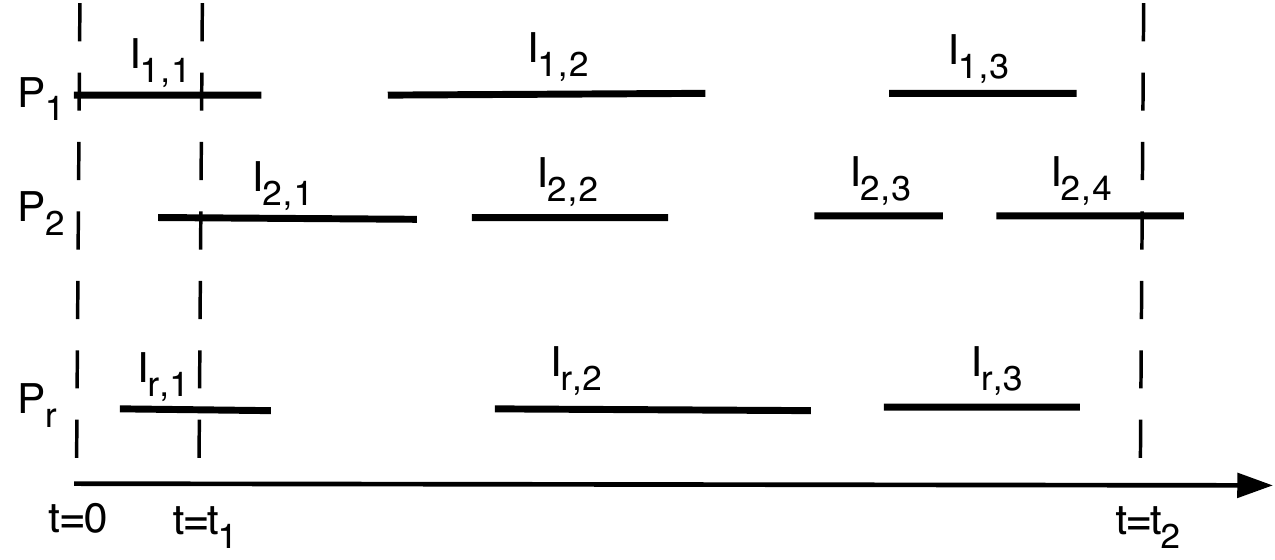}
\caption{Lifetime of $r$ Paths between a source-destination pair, and the corresponding time intervals when they are alive. $s$ and $d$ need to communicate between time $t=t_{1}$ and $t=t_{2}$}
\label{fig:pathLife}
\end{figure}

In Fig.~\ref{fig:interval} we showed how ``lifetime'' of a link (i.e., alive/dead) can be computed. Since paths comprise of links, a path between a source-destination node pair will also be alive/dead at different points of time. From the lifetime of links, we can compute the lifetime of paths in the following way. If the number of nodes (i.e., ANPs) in the network is $n$, there exists $n(n - 1)/2$ links with each having an individual lifetime. If a path $P$  is made up of links $l_1, l_2, \ldots, l_k$, the path $P$ is ``alive'' when all the links $l_1$ through $l_k$ are alive. Therefore, similar to Fig.~\ref{fig:interval} that shows the``lifetime'' of a link, we can construct a figure for ``lifetime'' of a path (Fig.~\ref{fig:pathLife}).
Once we have knowledge of lifetimes of paths, we can construct a route from the source node $s$ to the destination node $d$ with the fewest number of path switching in the following way.

The lifetimes of paths between a source-destination node pair $s$ and $d$ is shown in Fig.~ \ref{fig:pathLife}. The time intervals during which a path is alive is shown by solid lines in Fig.~ \ref{fig:pathLife}. We use the notation $P_j = \{I_{j, 1}, I_{j, 2}, \ldots, I_{j, j_k}\}$, to indicate that the path $P_j$ is alive during the $j_k$ time intervals $\{I_{j, 1}, I_{j, 2}, \ldots, I_{j, j_k}\}$, as shown in the Fig.~ \ref{fig:pathLife}. In the application scenario that we are considering, we want a communication channel to be open between the source node $s$ and the destination node $d$ for the entire duration of time from $t = t_1$ to $t = t_2$. Since it is possible that no single path between $s$ to $d$ remains alive for the entire duration from $t = t_1$ to $t = t_2$, a set of paths $\cal P$ may constitute a communication channel from $s$ to $d$ for the duration, where each path in $\cal P$ is alive only for a fraction of the time interval from $t = t_1$ to $t = t_2$.

Next we focus on the number of paths between $s$ and $d$ that we need to consider. Since the graph has $n$ nodes and $n(n -1)/2$ links, there could be as many as $(1 + (n - 2) + (n - 2)(n - 3) + \ldots + (n - 2)(n - 3)(n - 4) \ldots 2 = O(n^n))$ paths corresponding to 1-hop, 2-hop, $\ldots$, $(n - 1)$-hop paths between $s$ and $d$. It may also be noted that each of these paths will have a lifetime associated with it. Since examining $O(n^n)$ paths and their lifetimes will be too time consuming, we restrict our attention to only those paths between $s$ and $d$ whose number of hops is at most $k$, for some specified value of $k$. Restricting the number of hops in a path to at most $k$ (from $n -1$), we reduce the computational complexity from $O(n^n)$ to $O(n^k)$. Suppose the set of paths of at most $k$ hops is denoted at ${\cal P}_k$. The {\em minimum path switch routing} algorithm given below finds a subset ${\cal P}'_k \subseteq {\cal P}_k$ so that paths in ${\cal P}'_k$ maintain a communication channel between $s$ and $d$ for the entire time duration from $t = t_1$ to $t = t_2$ with the fewest number of path switchings.

\medskip
\noindent
{\em Minimum Path-Switch Routing Algorithm }

\vspace{0.1in}
\noindent
{\it Input:} The set ${\cal P}_k$ of paths between $s$ and $d$ with length at most $k$-hops, and associated lifetimes of each each path $P_i \in {\cal P}_k$, in the form of live intervals of $P_i$. If there are $r$ live intervals of $P_i$ in the time interval between $t = t_1$ to $t = t_1$, we denote $P_i = \lbrace I_{i, 1}, I_{i, 2}, \ldots, I_{i, r}\rbrace$.

\vspace{0.1in}
\noindent
{\it Output:} A subset ${\cal P}'_k \subseteq {\cal P}_k$ so that paths in ${\cal P}'_k$ maintain a communication channel between $s$ and $d$ for the entire time duration from $t = t_1$ to $t = t_2$ with the fewest number of path switchings.

\vspace{0.1in}
\noindent
{\it Comments:} The algorithm uses a greedy (locally optimum) approach to find the paths needed to  have one live path from $s$ to $d$ during the entire time interval $t = t_1$ and $t = t_2$. In Theorem 1, we prove that this locally optimum greedy approach indeed finds the globally optimal solution.

\vspace{0.1 in}
\begin{tabbing}
123\=123\=123\=123\=123\=123\=123\= \kill
1. {\em begin}\\
2. \> ${\cal P}'_k = \emptyset$;\\
3. \> $t_{start}~=~t_1$;\\
4. \> $t_{finish}~=~t_1$;\\
5. \>{\em While} ($t_{finish}~<~t_2$)~{\em do} \\
\> \> {\em begin}\\
\> \> \>  (i) {\em for all} $P_i \in {\cal P}_k $ {\em do}\\
\> \> \> \>  {\em if} ((start\_time $(I_{i,j}) \leq t_{start})$ \\
\> \> \> \>  {\em $\&\&$} $(t_{finish} < finish\_time(I_{i,j})$ for some $j$)\\
\> \> \> \> \>  {\em begin}\\
\> \> \>  \> \> \> (i) $t_{finish} = finish\_time(I_{i,j})$\\
\> \> \>  \> \> \>    (ii) ${\cal P}'_k = {\cal P}'_k \cup P_i$;\\
\> \> \> \> \>  {\em end}\\
\> \> \>  $t_{start}~=~t_{finish}$;\\
\> \> {\em  end}\\
6. {\em end}
\end{tabbing}

\begin{figure}
\centering
\includegraphics[width=0.48\textwidth, keepaspectratio]{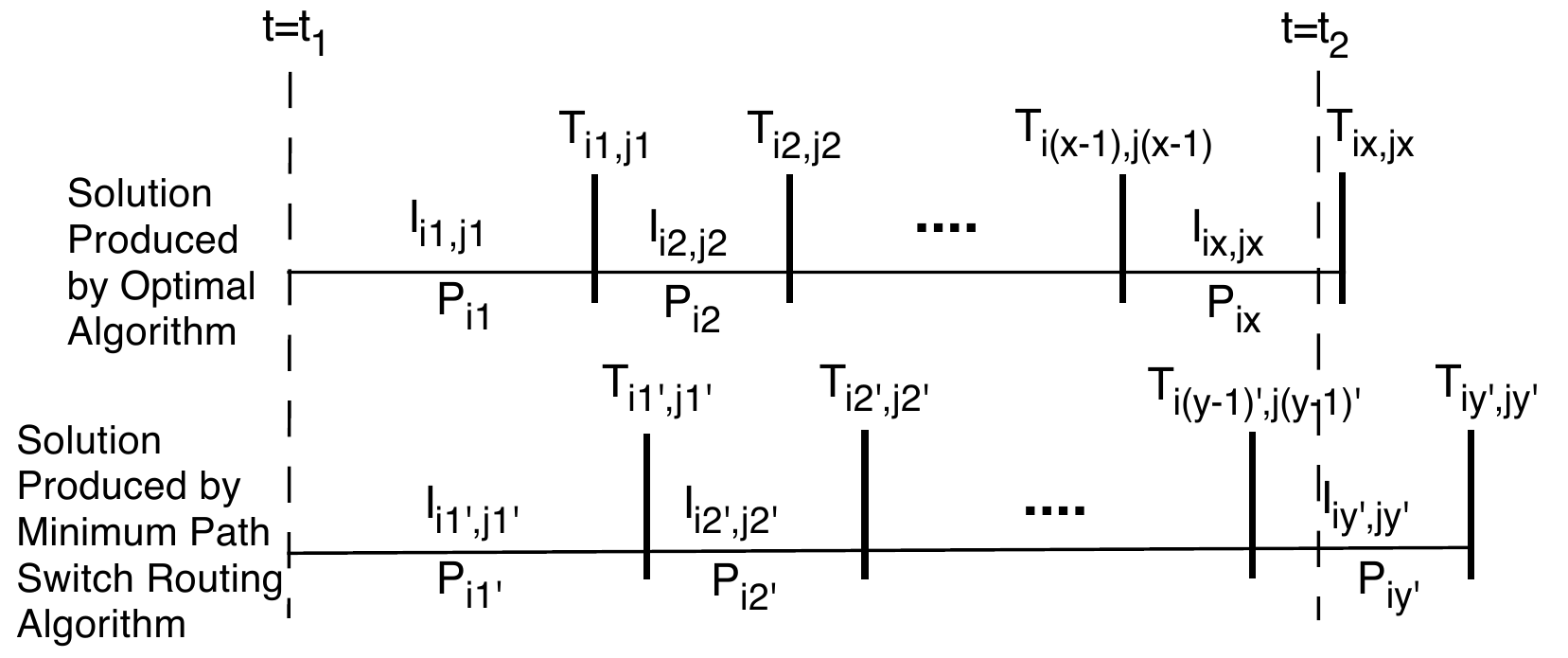}
\caption{Solution Produced by the Optimal Algorithm and the Minimum Path Switch Routing Algorithm}
\label{fig:proofdiagram}
\end{figure}

\vspace{0.1 in}
\noindent
{\em  Theorem 1:} The Minimum Path-Switch Routing Algorithm finds a set of paths so that a communication channel is open during the entire duration from $t = t_1$ to $t = t_1$ with the fewest number of path switches.

\vspace{0.1 in}
\noindent
{\em  Proof:} Suppose that an optimal algorithm selected the paths $\lbrace P_{i_1}, P_{i_2}, \ldots, P_{i_x}\rbrace$ and the Minimum Path-Switch Routingl algorithm selected the paths $\lbrace P_{i'_1}, P_{i'_2}, \ldots, P_{i'_y} \rbrace$, where $x < y$. The live intervals of the paths $\lbrace P_{i_1}, P_{i_2}, \ldots, P_{i_x}\rbrace$ that were selected by the optimal algorithm are $\lbrace I_{i_1, j_1}, I_{i_2, j_2}, \ldots, I_{i_x, j_x}\rbrace$ respectively. Similarly, the live intervals of the paths $\lbrace P_{i'_1}, P_{i'_2}, \ldots, P_{i'_x}\rbrace$ that were selected by the Minimum Path-Switch Routing algorithm are $\lbrace I_{i'_1, j'_1}, I_{i'_2, j'_2}, \ldots, I_{i'_y, j'_y}\rbrace$ respectively. The paths and intervals chosen by the two algorithms are shown in Fig.~\ref{fig:proofdiagram}. As shown in Fig.~\ref{fig:proofdiagram}, the finish times of the intervals $\lbrace I_{i_1, j_1}, I_{i_2, j_2}, \ldots, I_{i_x, j_x}\rbrace$ are denoted as $\lbrace T_{i_1, j_1}, T_{i_2, j_2}, \ldots, T_{i_x, j_x}\rbrace$ and the finish times of the intervals $\lbrace I_{i'_1, j'_1}, I_{i'_2, j'_2}, \ldots, I_{i'_y, j'_y}\rbrace$ are denoted as $\lbrace T_{i'_1, j'_1}, T_{i'_2, j'_2}, \ldots, T_{i'_y, j'_y}\rbrace$.

Since the Minimum Path-Switch Routing Algorithm chooses the path $P_i \in {\cal P}_k$ such that $P_i$ is live at $t_{start}$ and the finish time of the live interval containing $t_1$  is largest among the finish times of the intervals associated with all the paths, we can conclude that $T_{i'_1, j'_1} > T_{i_1, j_1}$. Therefore, replacing the path $P_{i_1}$ from the optimal soultion by the path $P_{i'_1}$ we will have a new optimal solution $\lbrace P_{i'_1}, P_{i_2}, \ldots, P_{i_x}\rbrace$. Because of nature of the path selection criteria of the Minimum Path-Switch Routing Algorithm, we can conclude that $T_{i'_2, j'_2} > T_{i_2, j_2}$. Therefore, replacing the path $P_{i_2}$ from the new optimal soultion by the path $P_{i'_2}$ we will have yet another optimal solution $\lbrace P_{i'_1}, P_{i'_2}, \ldots, P_{i_x}\rbrace$. Continuing this process, we can  get an optimal solution  $\lbrace P_{i'_1}, P_{i_2}, \ldots, P_{i'_x}\rbrace$. This implies that the Minimum Path-Switch Routing Algorithm will select only $x$ paths instead of $y$, ($x < y$) to have an open communication channel for the enire duration of $t = t_1$ to $t = t_2$, and since $x$ is the optimal number of paths needed for this purpose, the Minimum Path-Switch Routing Algorithm produces an optimal solution.

\section{Design for Coverage - Problem Formulation}
\label{sec:coverageProbFormulation}

In this section, we discuss the coverage model of the network formed by the ANPs. As shown in Fig.~\ref{fig:airCorridor1}, an air corridor through which the combat aircrafts fly towards their destination can be modeled as a collection of rectangular parallelepipeds. As the combat aircrafts must have access to the AN as they fly through the air corridor, all points inside the air-corridor must have radio coverage  at all times. As the shape of an air corridor can be quite complex, we view that the complex shape can be approximated with rectangular parallelepipeds as shown in Fig.~\ref{fig:airCorridor2}.

The length, width and height of this section of the air-corridor are denoted as $L_{ac}$, $W_{ac}$ and $H_{ac}$, respectively. The radius of the circular orbit of the ANPs is denoted by  $r_{o}$ and the number of ANPs in each such orbit is denoted by $n$. We assume that the ANPs move around in their orbit with uniform velocities. The coverage volume of each ANP is defined as a spherical volume of radius $r_{s}$ with the ANP being at the center of the sphere. We assume that the orbits of the ANPs are located at the top surface of the air corridor so that they do not cause any  hindrance in the flight path of the combat aircrafts.  This is shown in Fig.~\ref{fig:airCorridor4}, where  two circular orbits, each of them containing $5$ ANPs, are located at the top surface of the air corridor. The number of orbits located at the top surface of the air corridor section is denoted as $m$, and accordingly the total number of ANPs in the network is given by $mn$. The goal of the coverage problem  is to provide complete coverage  at all times of the entire air corridor with the fewest number of ANPs. In this problem, a complete coverage must be provided irrespective of the locations of the ANPs as they move continuously in their respective orbits.

\begin{figure*}[!t]
\centering
\subfigure[$5$ ANPs moving in a circular orbit]{\includegraphics[width=0.35\textwidth, keepaspectratio]{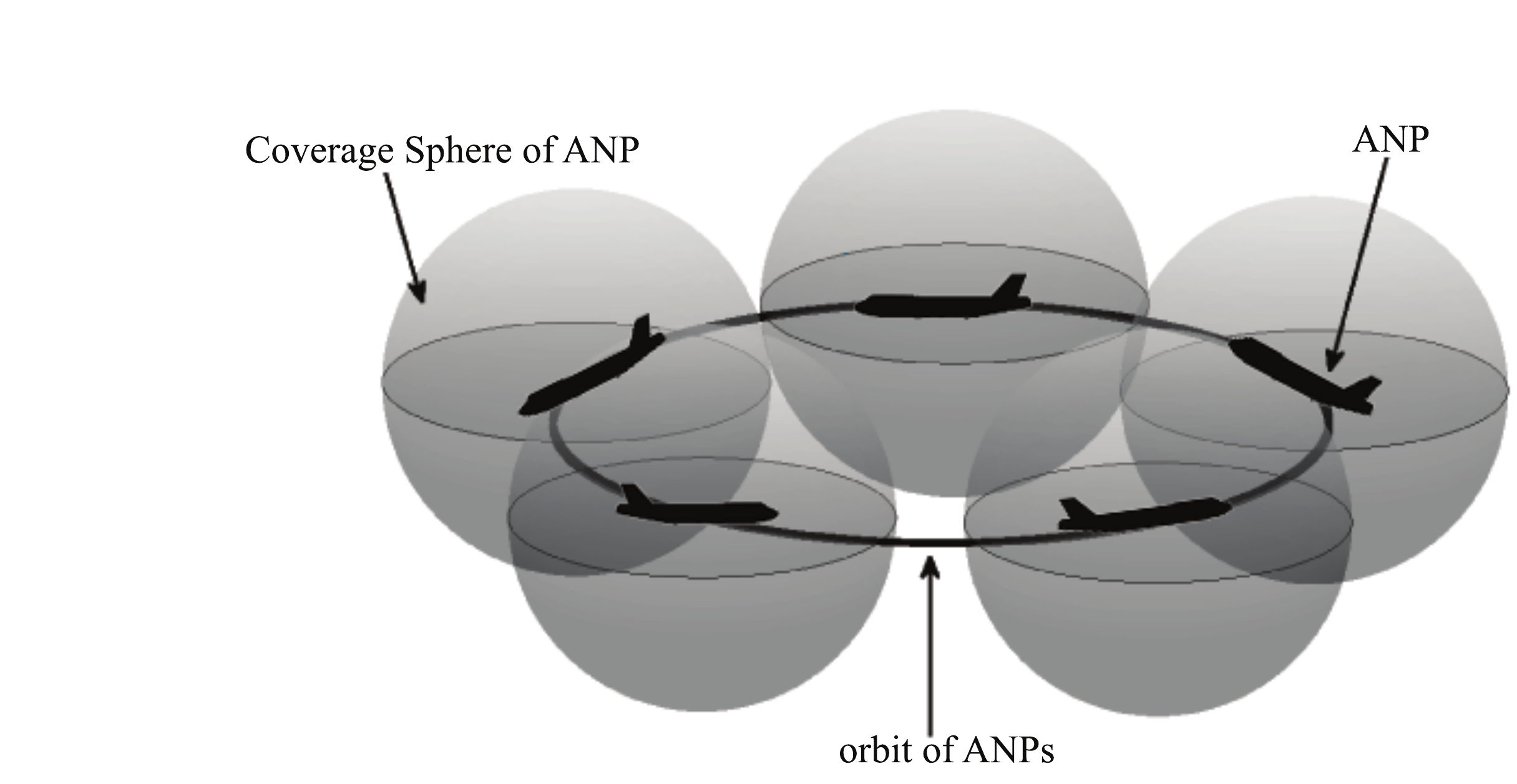}\label{fig:fig1}}
\hfill
\subfigure[Top View of the coverage spheres and circular orbit of $5$ ANPs]{\includegraphics[width=0.3\textwidth, keepaspectratio]{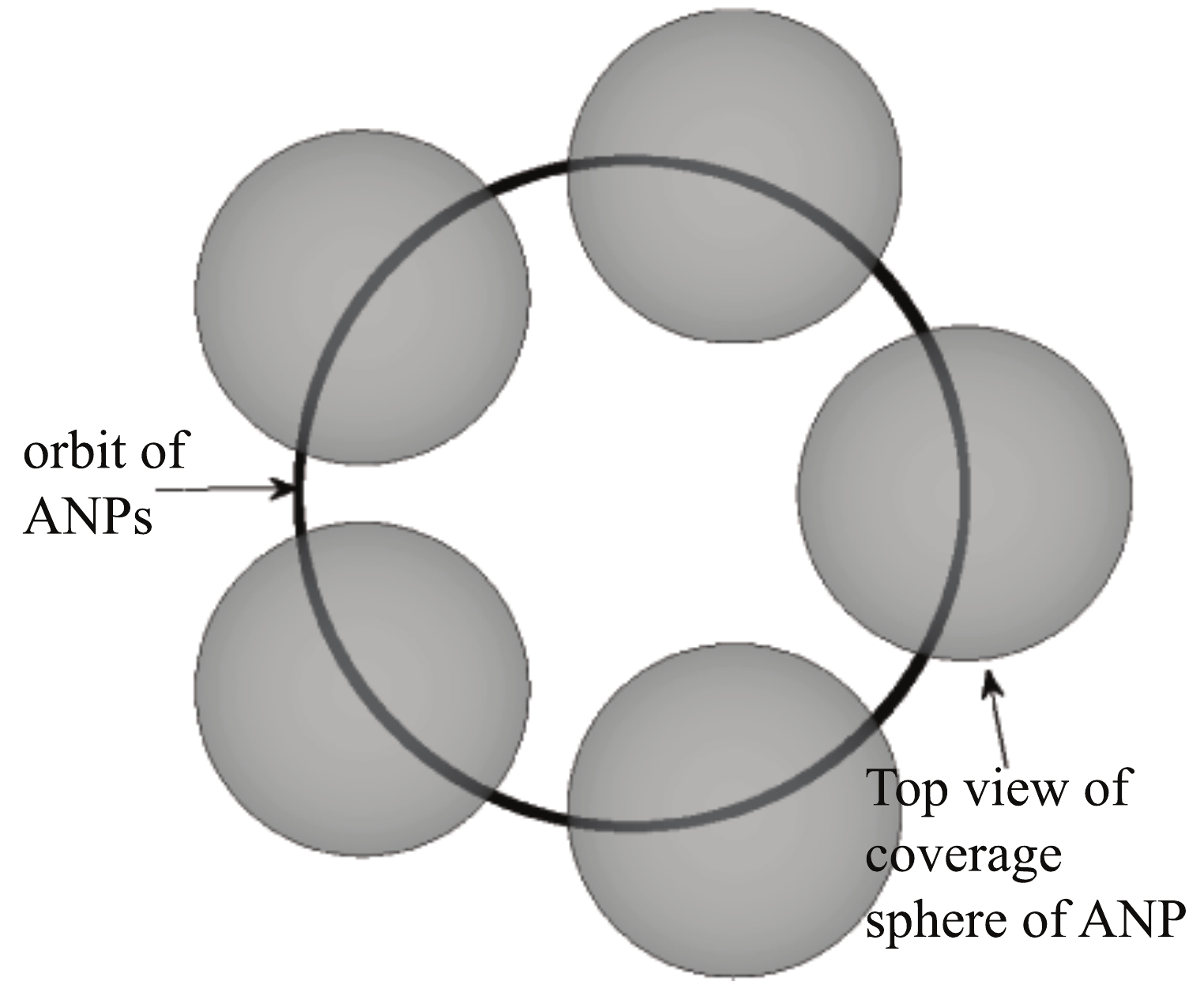}\label{fig:fig2}}
\hfill
\subfigure[Front View of the coverage spheres and circular orbit of $5$ ANPs ]{\includegraphics[width=0.25\textwidth, keepaspectratio]{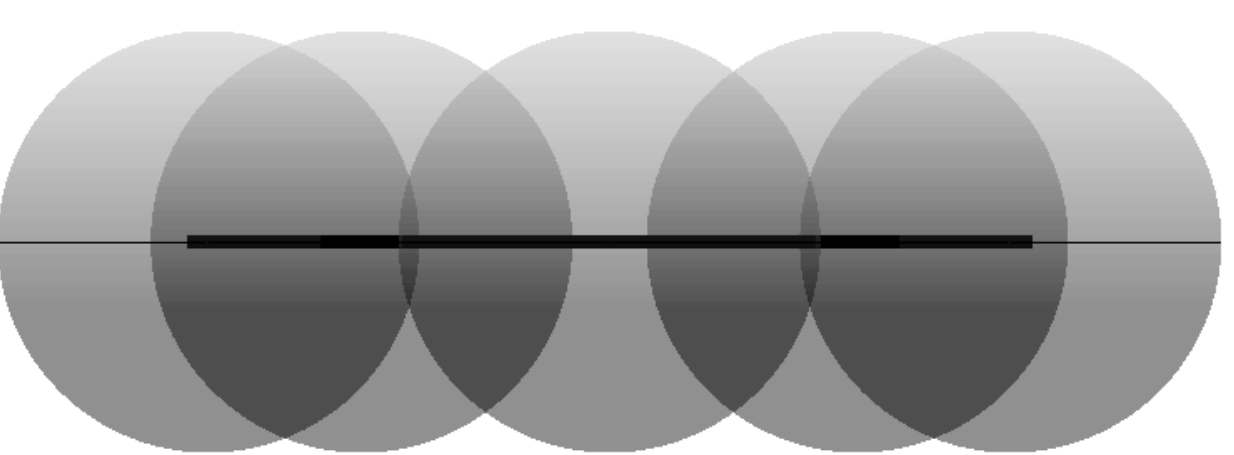}\label{fig:fig3}}
\caption{Case I: orbit radius of ANPs $(r_{o}) >$ radius of coverage sphere $(r_{s})$}
\label{fig:fig123}
\end{figure*}

\begin{figure*}[!t]
\centering
\subfigure[$5$ ANPs moving in a circular orbit]{\includegraphics[width=0.35\textwidth, keepaspectratio]{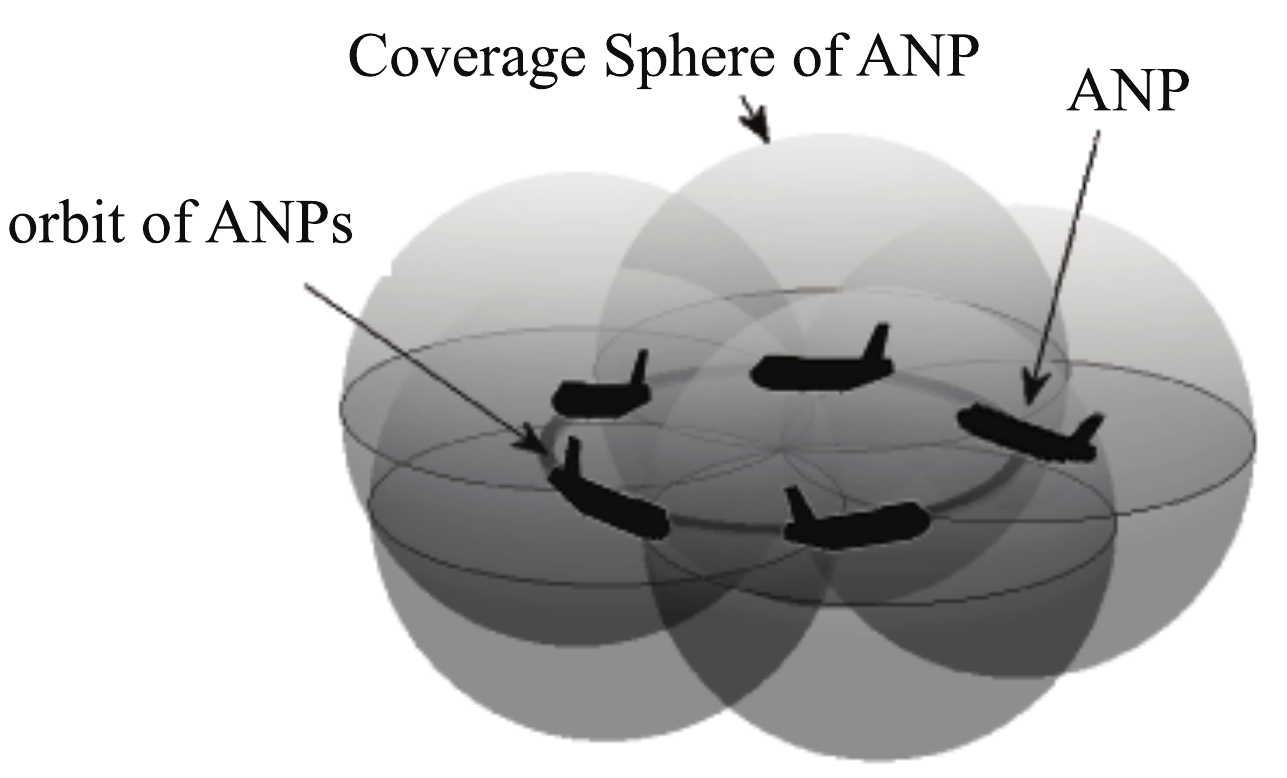}\label{fig:fig4}}
\hfill
\subfigure[Top View of the coverage spheres and circular orbit of $5$ ANPs]{\includegraphics[width=0.3\textwidth, keepaspectratio]{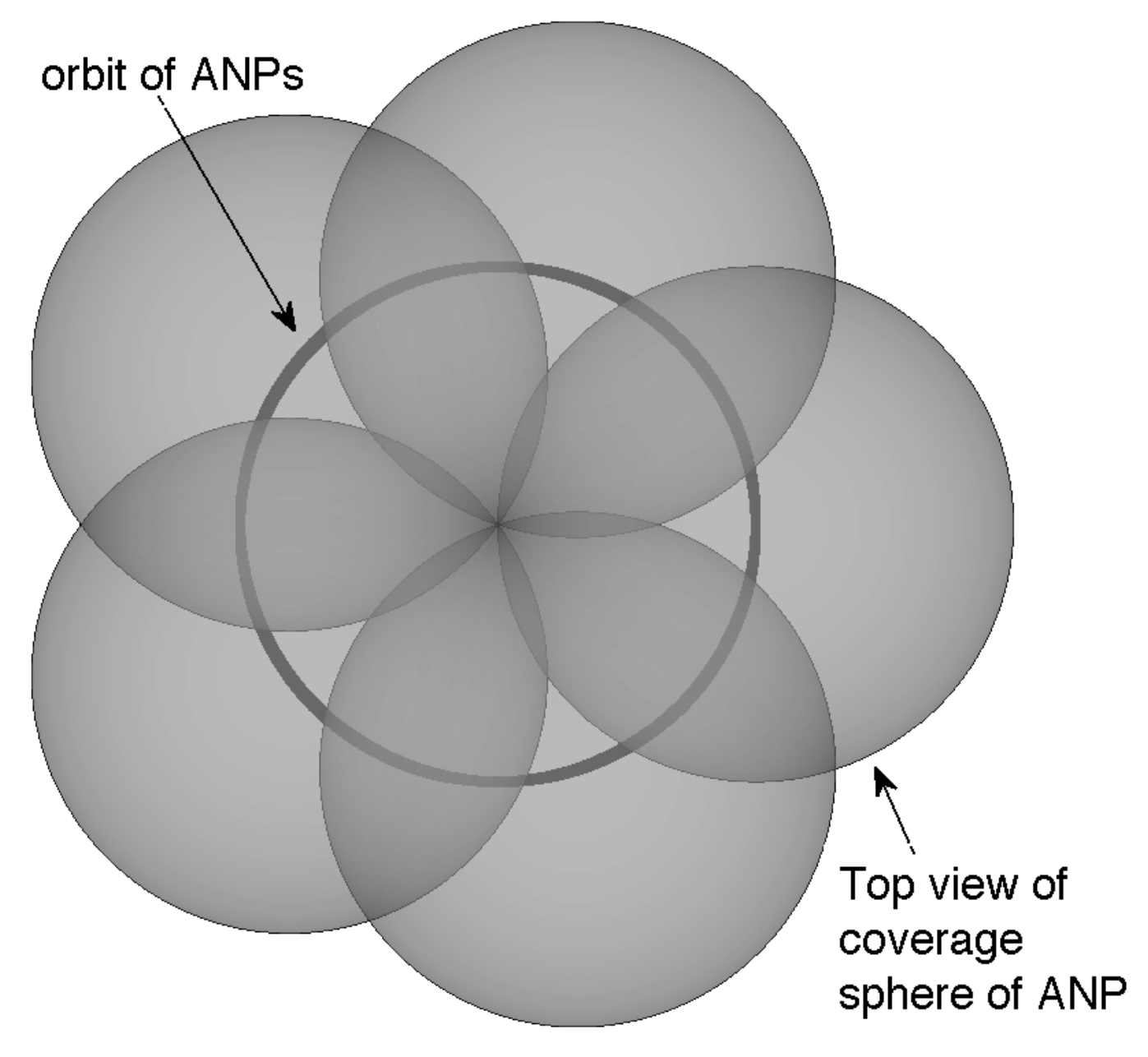}\label{fig:fig5}}
\hfill
\subfigure[Front View of the coverage spheres and circular orbit of $5$ ANPs ]{\includegraphics[width=0.25\textwidth, keepaspectratio]{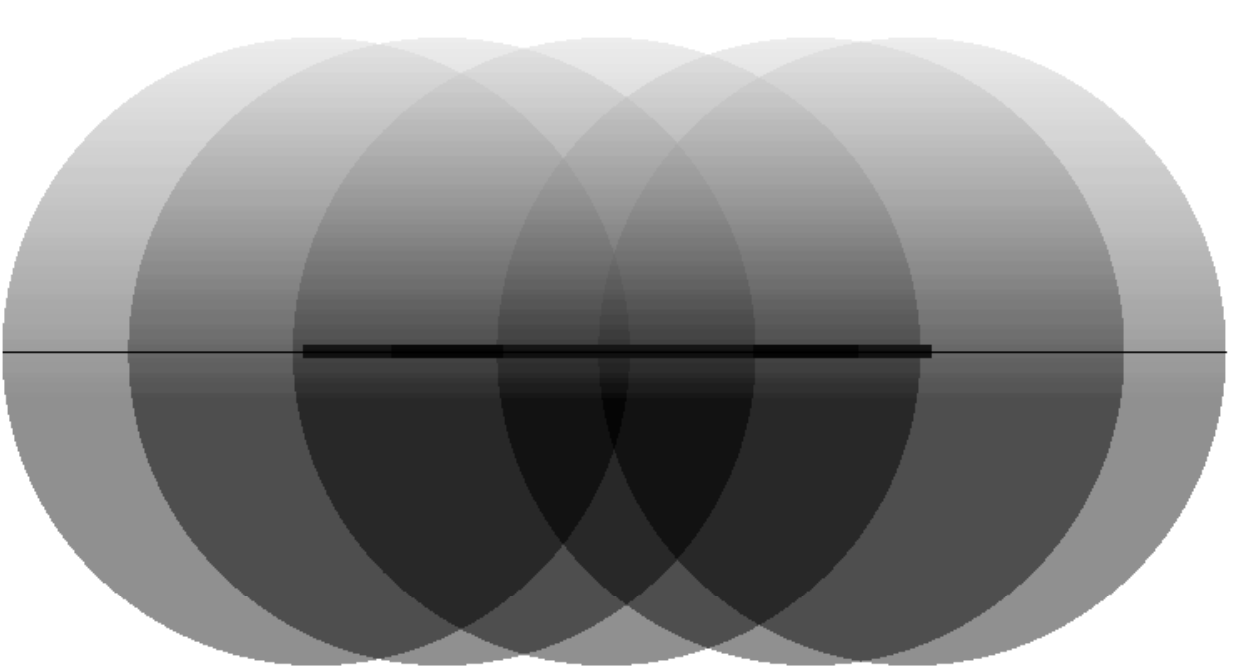}\label{fig:fig6}}
\caption{Case II: orbit radius of ANPs $(r_{o}) = $ radius of coverage sphere $(r_{s})$}
\label{fig:fig456}
\end{figure*}

\begin{figure*}[!t]
\centering
\subfigure[$5$ ANPs moving in a circular orbit]{\includegraphics[width=0.35\textwidth, keepaspectratio]{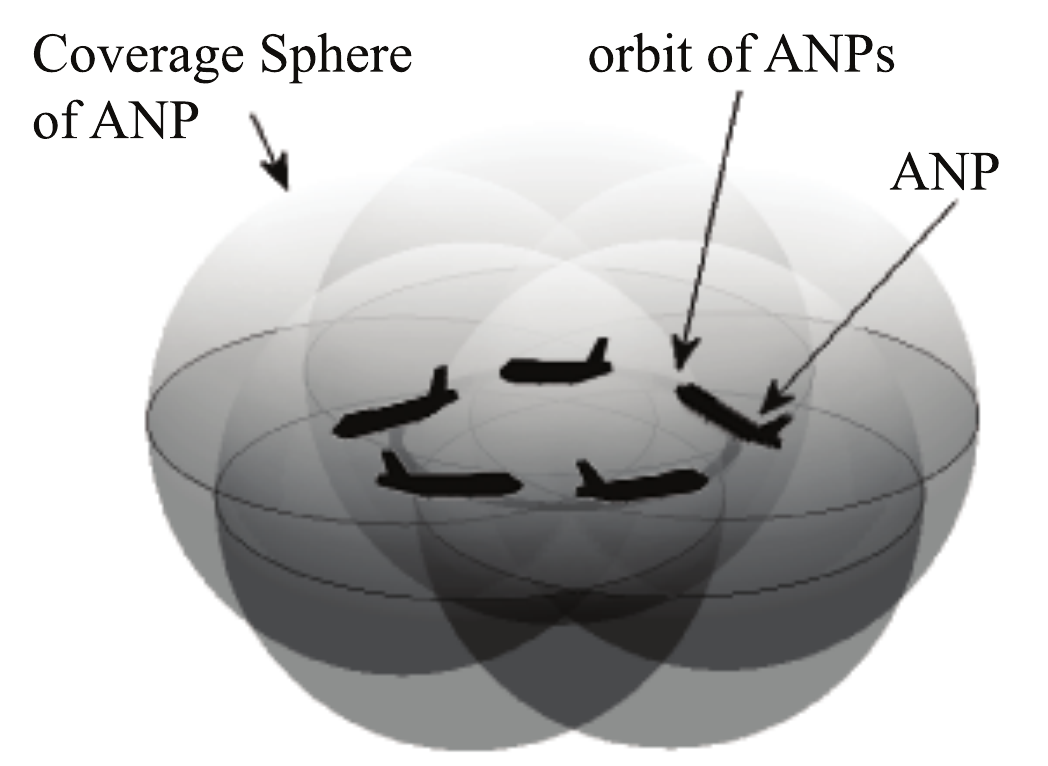}\label{fig:fig7}}
\hfill
\subfigure[Top View of the coverage spheres and circular orbit of $5$ ANPs]{\includegraphics[width=0.3\textwidth, keepaspectratio]{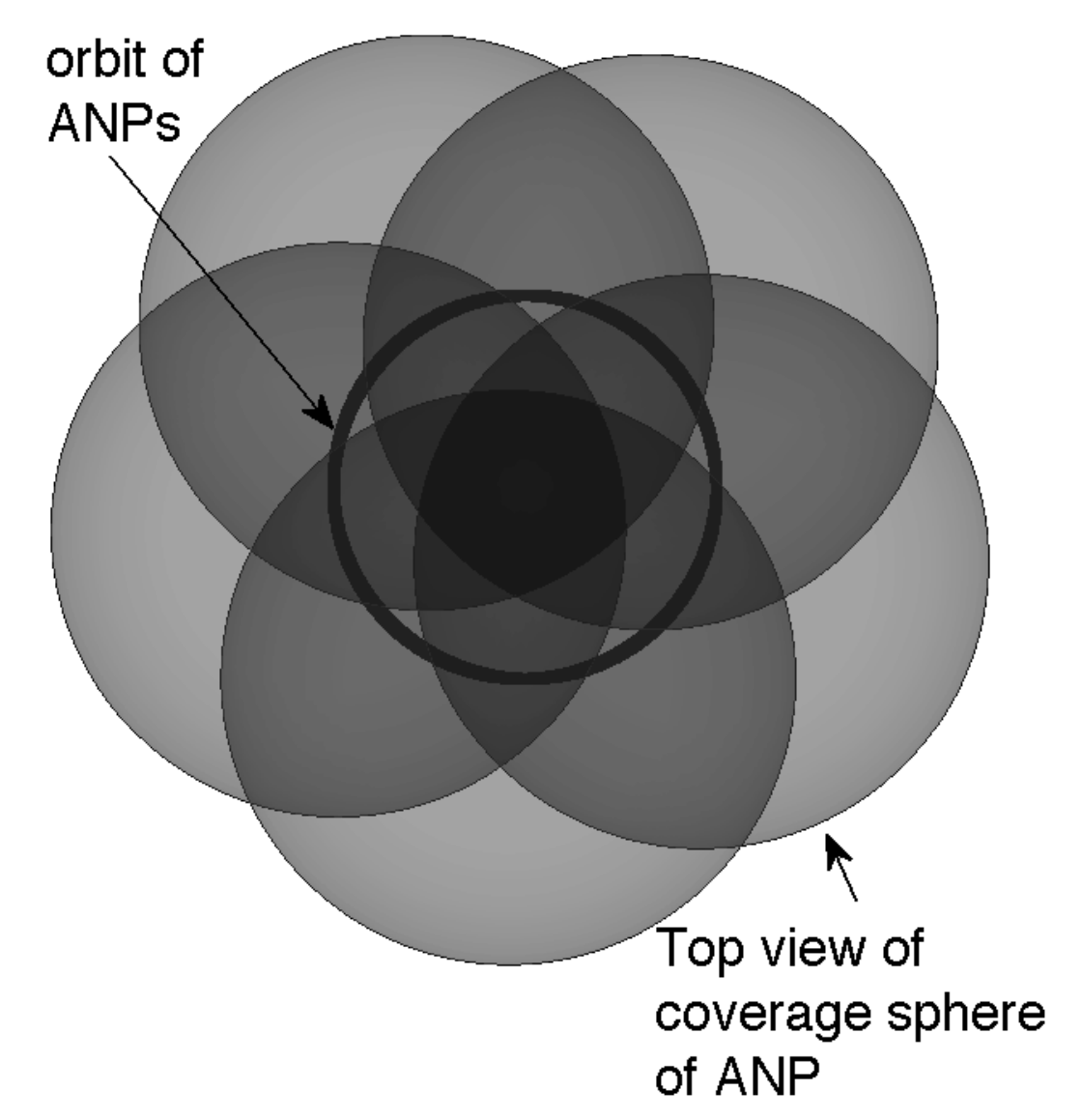}\label{fig:fig8}}
\hfill
\subfigure[Front View of the coverage spheres and circular orbit of $5$ ANPs ]{\includegraphics[width=0.25\textwidth, keepaspectratio]{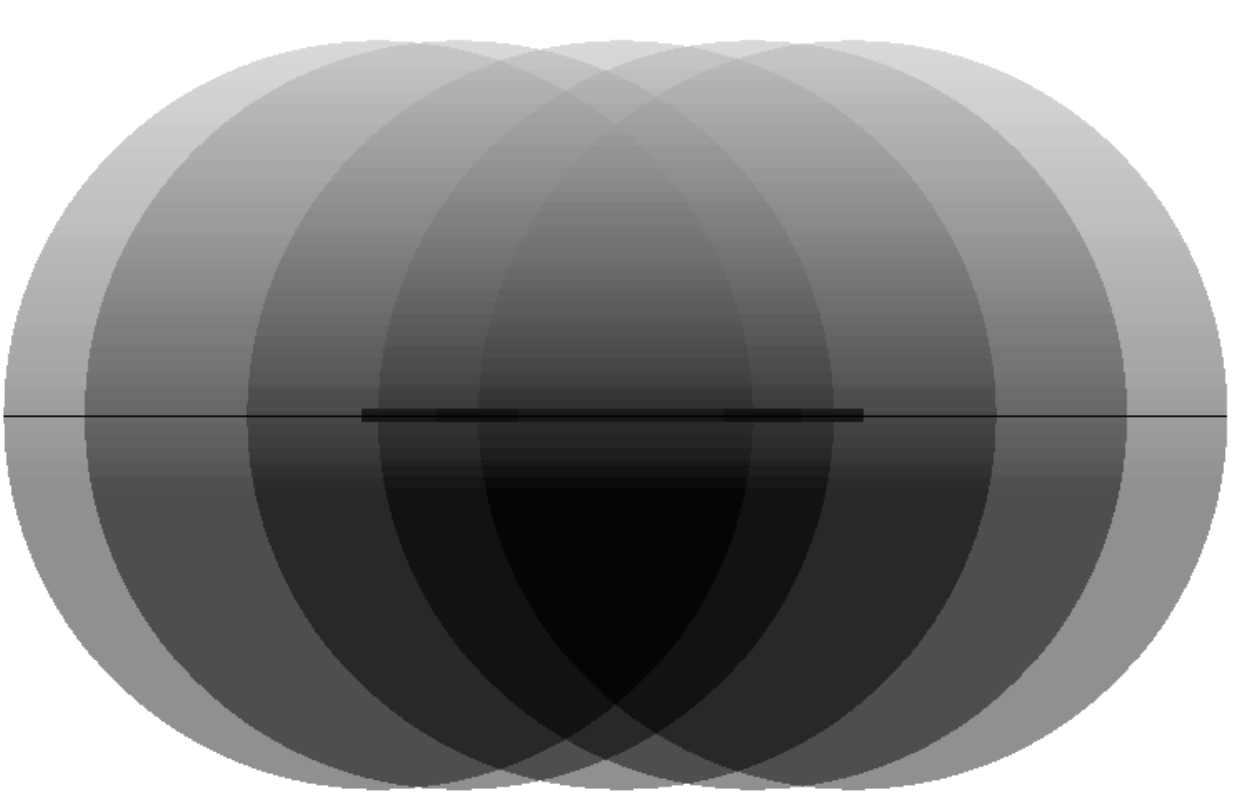}\label{fig:fig9}}
\caption{Case II: orbit radius of ANPs $(r_{o}) < $ radius of coverage sphere $(r_{s})$}
\label{fig:fig789}
\end{figure*}

In Figs.~ \ref{fig:fig123}, \ref{fig:fig456}, \ref{fig:fig789}, we can see the 3D view, top-view and front-view of $5$ ANPs in circular orbits and the volume covered by them for three different cases. In case I (Fig.~\ref{fig:fig123}), the value of orbit radius ($r_{o}$) is greater than that of the spherical coverage volume ($r_{s}$) of each ANP. It can be clearly seen from Fig.~\ref{fig:fig2}, that there is an open space inside the orbit, that is not covered by any of the ANPs as they move in the orbit. To increase the intersection volume, $r_{o}$ can be at most $r_{s}$ (case II). In Fig.~\ref{fig:fig456}, $r_{o}=r_{s}$, and we can see from Fig.~\ref{fig:fig5}, that the spheres meet at a single point inside the orbit. We call the intersection of adjacent spheres as {\em leaf}, the top view of which is visible in the figure. To increase the intersection of the spheres further, we need to decrease $r_{o}$ even more, and thus bringing in the ANPs even more closer to each other. This is case III (shown in Fig.~\ref{fig:fig789}), where $r_{o} < r_{s}$.  

The coverage problem can be stated as follows: Given the rectangular parallelepipeds in terms of $L_{ac}$, $W_{ac}$ and $H_{ac}$ and the radius of the coverage sphere associated with an ANP,  $r_{s}$, find the radius of the orbit of the ANPs $r_{o}$, and the number of ANPs in each orbit ($n$), the entire volume of the air corridor is covered at all times with the fewest number of ANPs.

\begin{center}
\begin{figure*}[tbh]
    \begin{minipage}[tbh]{0.35\linewidth}
        \centering
       \includegraphics[width=0.8\textwidth, keepaspectratio]{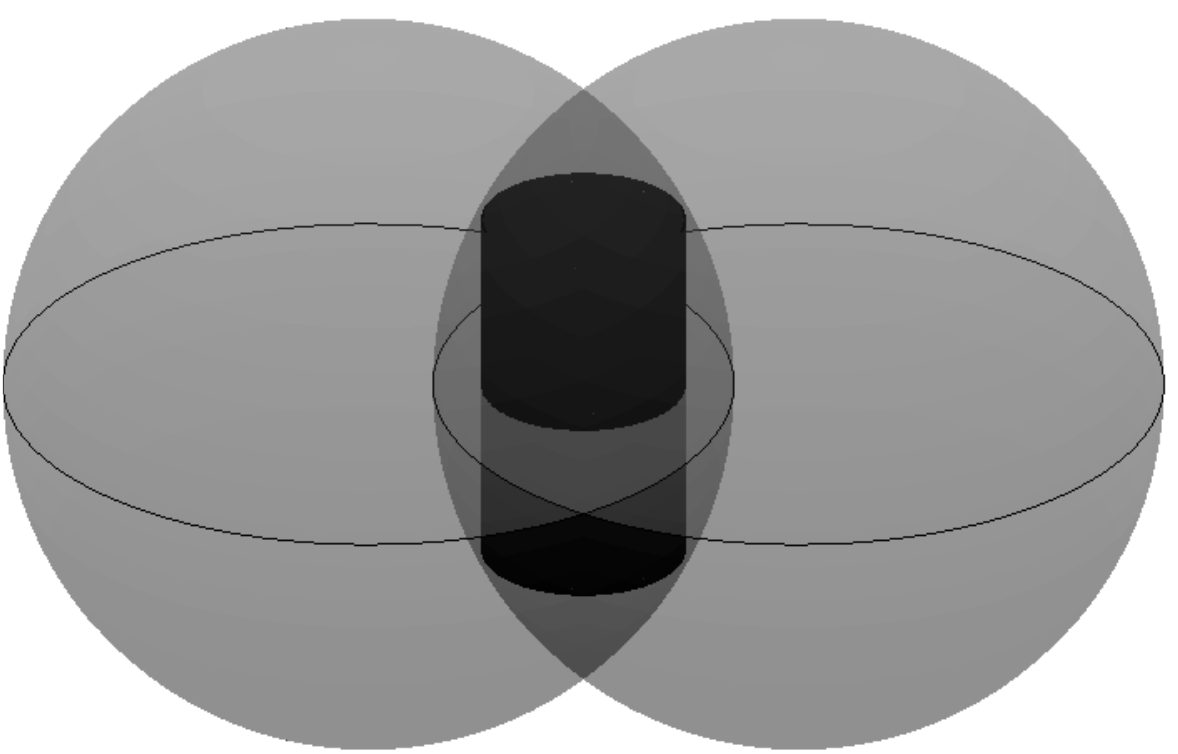}
\caption{Intersection volume of two spheres; cylindrical volume cut from the intersection volume}
\label{fig:int2s}
    \end{minipage}
    \hfill
    \begin{minipage}[tbh]{0.6\linewidth}
        \centering
       	\subfigure[3-D view]{\includegraphics[width=0.4\textwidth, keepaspectratio]{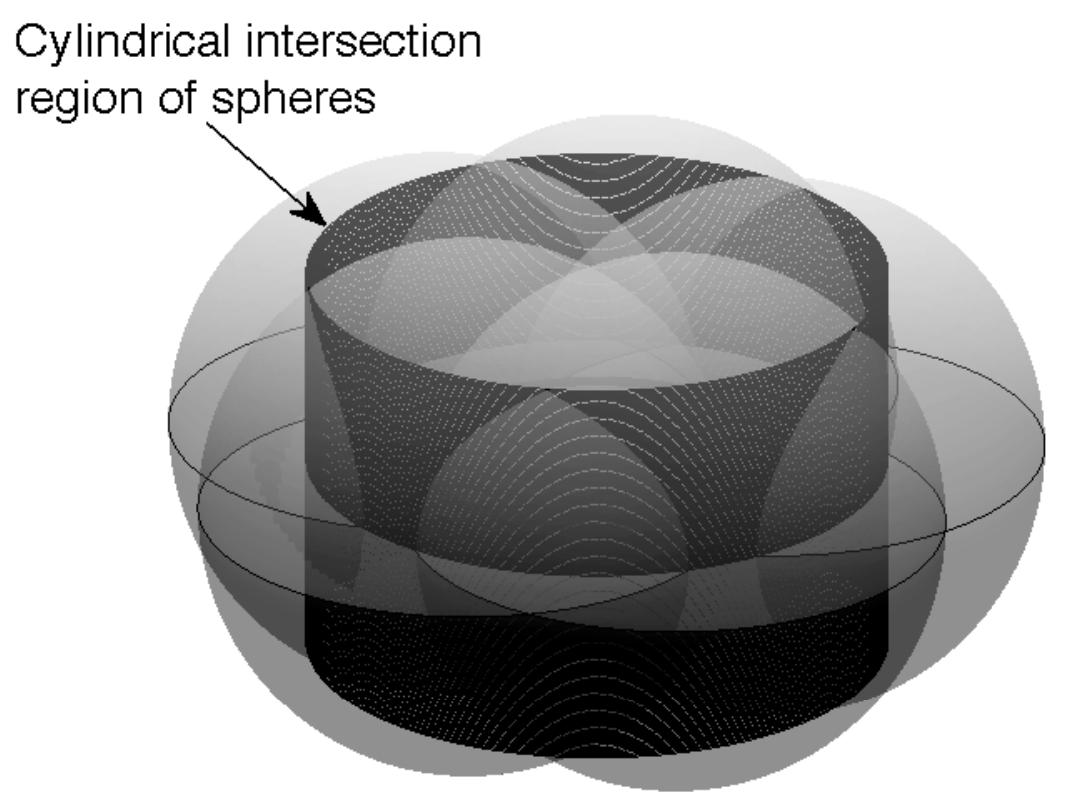}\label{fig:fig10}}
\hfill
\subfigure[Front View]{\includegraphics[width=0.4\textwidth, keepaspectratio]{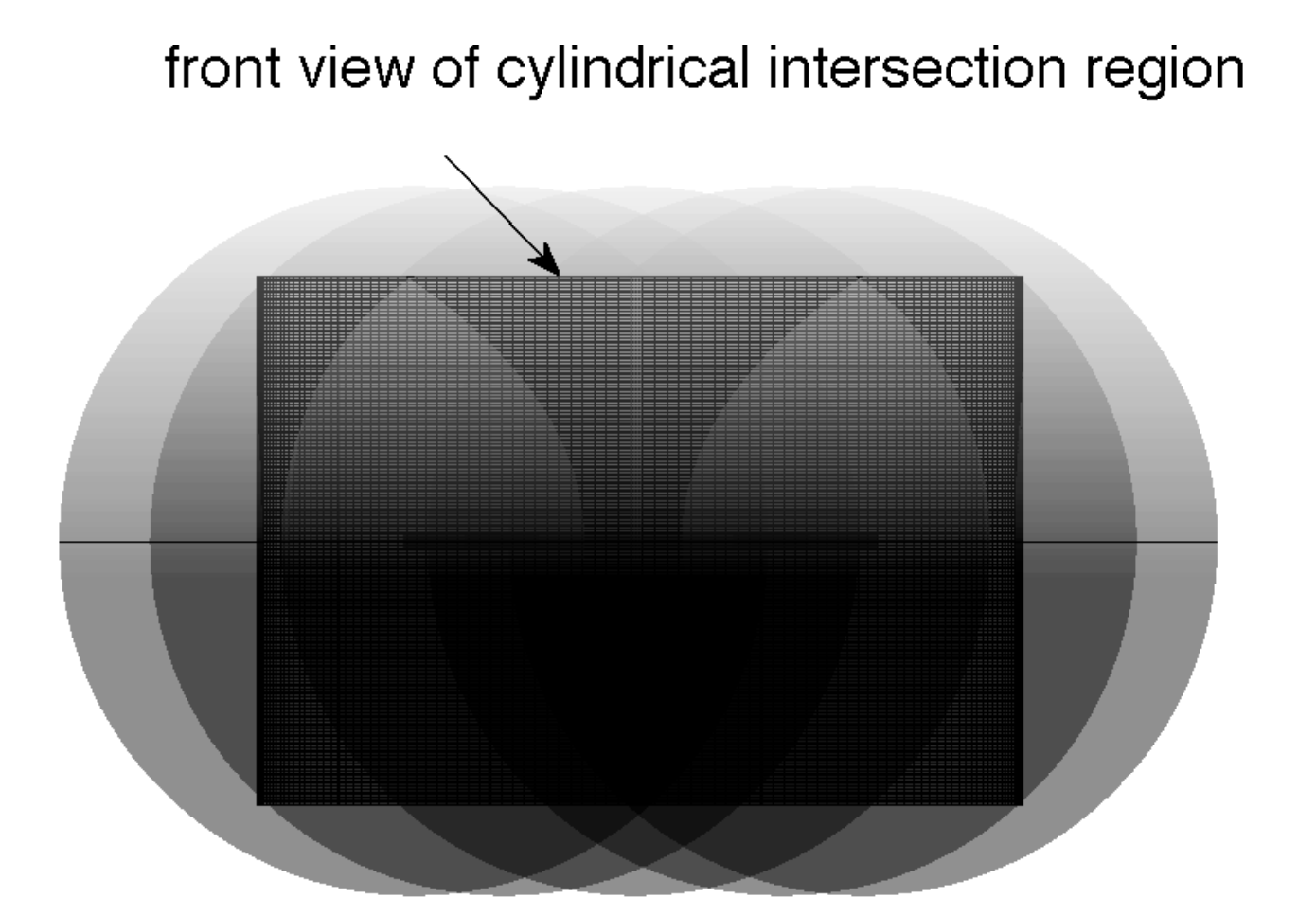}\label{fig:fig12}}
\caption{Cylindrical volume cut from the intersection of the coverage spheres of ANPs ($r_{s}>r_{o}$)}
\label{fig:fig101112}
    \end{minipage}
\end{figure*}
\end{center}

Intersection of coverage spheres of the ANPs create a coverage volume. For two intersecting spheres, the intersection volume is shown in Fig.~\ref{fig:int2s}. As the ANPs move in their orbits, the associated coverage spheres move with them and consequently the volume that is covered by the moving ANPs also changes. As a consequence some volume will be covered only a part of the time. However, a part of the intersection volume will be covered at all times irrespective of the positions of the ANPs as they move in their orbits. This is defined as the {\em invariant coverage volume}. We would like to use this {\em invariant coverage volume} as building blocks in order to fill up the air corridor modeled in the form of a rectangular parallelopiped. Since the {\em invariant coverage volume} is irregular-shaped, it is difficult to use it as a building block. For ease of coverage using a  building block with a regular shape, we extract a cylindrical volume out of this invariant volume  and use it to fill up the rectangular parallelopiped. Such a cylinder is shown in Fig.~\ref{fig:int2s}. Different views of such a cylindrical section for $5$ intersecting ANPs in a circular orbit are shown in Fig.~\ref{fig:fig101112}.

\begin{figure}[!t]
\centering
\includegraphics[width=0.35\textwidth, keepaspectratio]{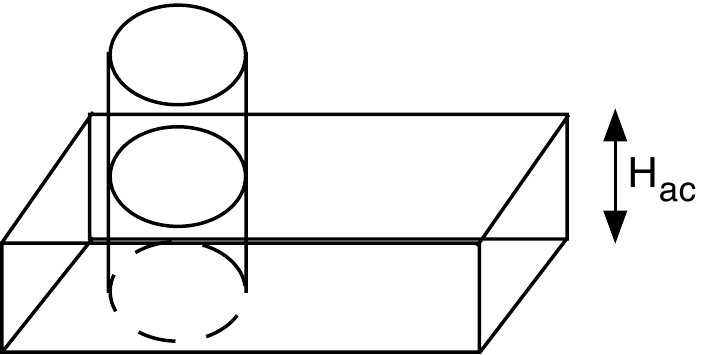}
\caption{Air-corridor being filled up with cylindrical sections - the cylinder shown is formed due to the ANPs in a particular orbit on the top surface}
\label{fig:airCorridor3}
\end{figure}

As we have decided to use a cylinder as the building block to cover the air corridor, we need to know the 
height and radius of the circular surface of such cylindrical blocks, denoted by $2h_{c}$ and $r_{c}$, respectively. The height and radius of the invariant coverage cylinder are determined by (i) the orbit radius ($r_{o}$), (ii) the number of ANPs per orbit ($n$) and (iii) the radius of the spherical coverage volume of each ANP ($r_{s}$). As mentioned earlier, in this design the ANPs and their orbits are placed on the top surface of the air corridor. As a consequence, the top half of the invariant coverage cylinder cannot be utilized  and only the bottom half of the cylindrical volume (of height $h_{c}$) will be used for the coverage of the rectangular parallelopiped. Therefore, in order to cover the height of air corridor, one must satisfy the constraint $h_{c}\geq H_{ac}$ (Fig.~\ref{fig:airCorridor3}). Once this constraint is satisfied, the problem reduces to cover the plane defined by $L_{ac} \times W_{ac}$ with circles of radius $r_{c}$ with a goal to minimize the total number of ANPs required ($mn$).

\begin{figure}[!t]
\centering
\includegraphics{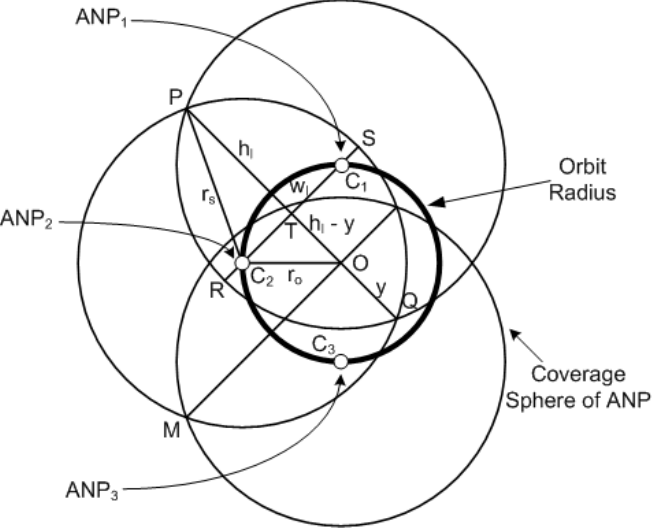}
\caption{Top View of Three Consecutive Intersecting Spheres Moving in a Orbit}
\label{fig:top}
\end{figure}

We investigate the structure of the coverage volume. In Fig.~\ref{fig:top}, we have shown the top view through the center of orbit of three consecutive spheres (out of total $n$ of them) intersecting with each other and moving around in the circular orbit with center at $O$. The center of the spheres are denoted as $C_{1}$, $C_{2}$ and $C_{3}$. The radius of the circular orbit is $\overline{OC_{1}} = \overline{OC_{2}} = \overline{OC_{3}} = r_{o}$. All the spheres are of uniform radius $r_{s}$, and moving in a uniform velocity. The term {\em leaf} is used to refer to the intersection between two adjacent spheres. In the top view, it can be seen as the intersection of two circular arcs $\wideparen{PSQ}$ and $\wideparen{QRP}$.  The length of the {\em leaf} $\overline{PQ}$ is denoted by $2h_{l}$ ($\overline{PT} = \overline{TQ} = h_{l}$). Distance from the center of the orbit center $O$ to the end point of the {\em leaf} $Q$ is denoted by $y$. Therefore,
\[
	\overline{TO} = \overline{TQ} - \overline{OQ} = h_{l} - y
\]
The width of the {\em leaf} ($\overline{TS}$) is denoted by $w_{l}$. Angle between two adjacent {\em leaves} $\angle{POM}$ is given by $\theta$. For $n$ number of spheres moving in the orbit, $\theta$ is given by $\frac{2\pi}{n}$. Therefore,
\[
	\angle{POC_{2}} = \angle{C_{2}OM} = \frac{\theta}{2} = \frac{\pi}{n}
\]

\begin{figure}[!t]
\centering
\includegraphics{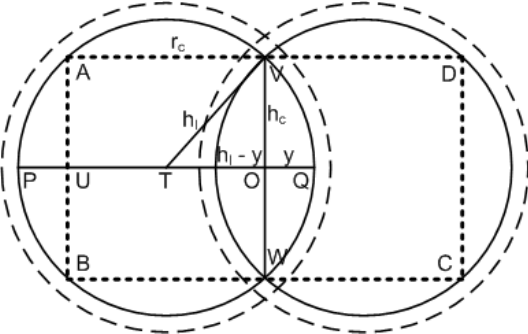}
\caption{Side View of Intersecting Spheres and the Leaves}
\label{fig:side}
\end{figure}

Now, in Fig.~ \ref{fig:side}, the side view of two intersecting spheres and leafs are shown. The intersecting spheres are shown using dashed line, whereas the intersecting {\em leaves} are shown using solid lines. We cut the largest cylindrical volume from the intersecting region such that this is covered by at least one sphere at all times, as the spheres move around in the orbit. From Fig.~ \ref{fig:top}:
\begin{eqnarray}
\label{eq:hlwl}
\overline{PT}^{2} & = & \overline{PC_{2}}^{2} - \overline{C_{2}T}^{2} \nonumber\\
i.e., h_{l}^{2} & = & r_{s}^{2} - (\overline{C_{2}S} - \overline{TS})^{2} \nonumber\\
i.e., h_{l}^{2} & = & r_{s}^{2} - (r_{s}-w_{l})^{2} \nonumber\\
h_{l} & = & \sqrt{w_{l}(2r_{s}-w_{l})}
\end{eqnarray}
Also, from $\triangle{C_{2}TO}$:
\begin{eqnarray}
\label{eq:rswl}
\overline{C_{2}T} & = & \overline{TO} tan\frac{\theta}{2} \nonumber\\
r_{s} - w_{l} & = & (h_{l}-y) tan\frac{\pi}{n}\\
\label{eq:rsro}
\overline{C_{2}T} & = & \overline{C_{2}O} sin\frac{\theta}{2} \nonumber \\
r_{s} - w_{l} & = & r_{o}sin\frac{\theta}{2} \nonumber \\
w_{l} & = & r_{s} - r_{o}sin\frac{\pi}{n} 
\end{eqnarray}


From equations \ref{eq:hlwl}, \ref{eq:rswl} and \ref{eq:rsro}, it is clear that all variables $h_{l}$, $w_{l}$ and $y$ can be represented through variables $r_{o}$ and $n$ only. Now from Fig. \ref{fig:side}, it is clear that the cylindrical volume with largest height that remains covered during the movement of the spheres, has radius of the circular surface ($r_{c}$) equal to length $\overline{VA}$. This can be calculated as :
\begin{eqnarray}
\label{eq:rcy}
\overline{VA} & = & \overline{OU} \nonumber\\
& = & 2 (\overline{TQ} - \overline{OQ}) \nonumber\\
i.e., r_{c} & = & 2(h_{l}-y)
\end{eqnarray}
The height of the cylinder is given by $\overline{VW} = 2\overline{VO} = 2 h_{c}$, calculated as:
\begin{eqnarray}
\label{eq:hcy}
\overline{VO}^{2} & = & \overline{VT}^{2} - \overline{TO}^{2}\nonumber \\
h_{c}^{2} & = & h_{l}^{2} - (h_{l}-y)^{2} \nonumber\\
h_{c} & = & \sqrt{y(2h_{l}-y)}
\end{eqnarray}

\begin{figure*}[!t]
\centering
\subfigure[Radius of Cylinder]{\includegraphics[width=0.325\textwidth,keepaspectratio]{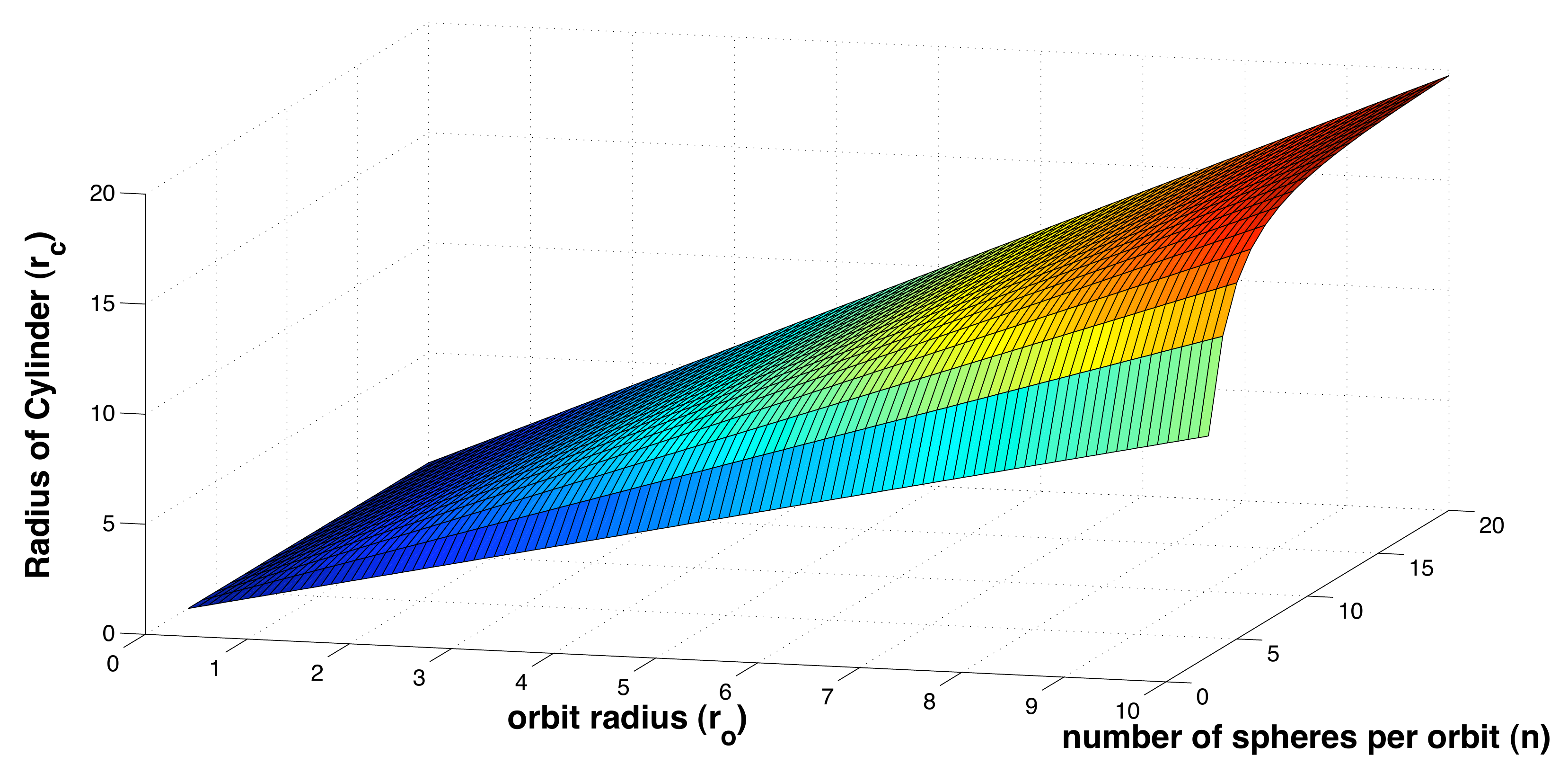}\label{fig:radius}}
\hfill
\subfigure[Height of Cylinder]{\includegraphics[width=0.325\textwidth,keepaspectratio]{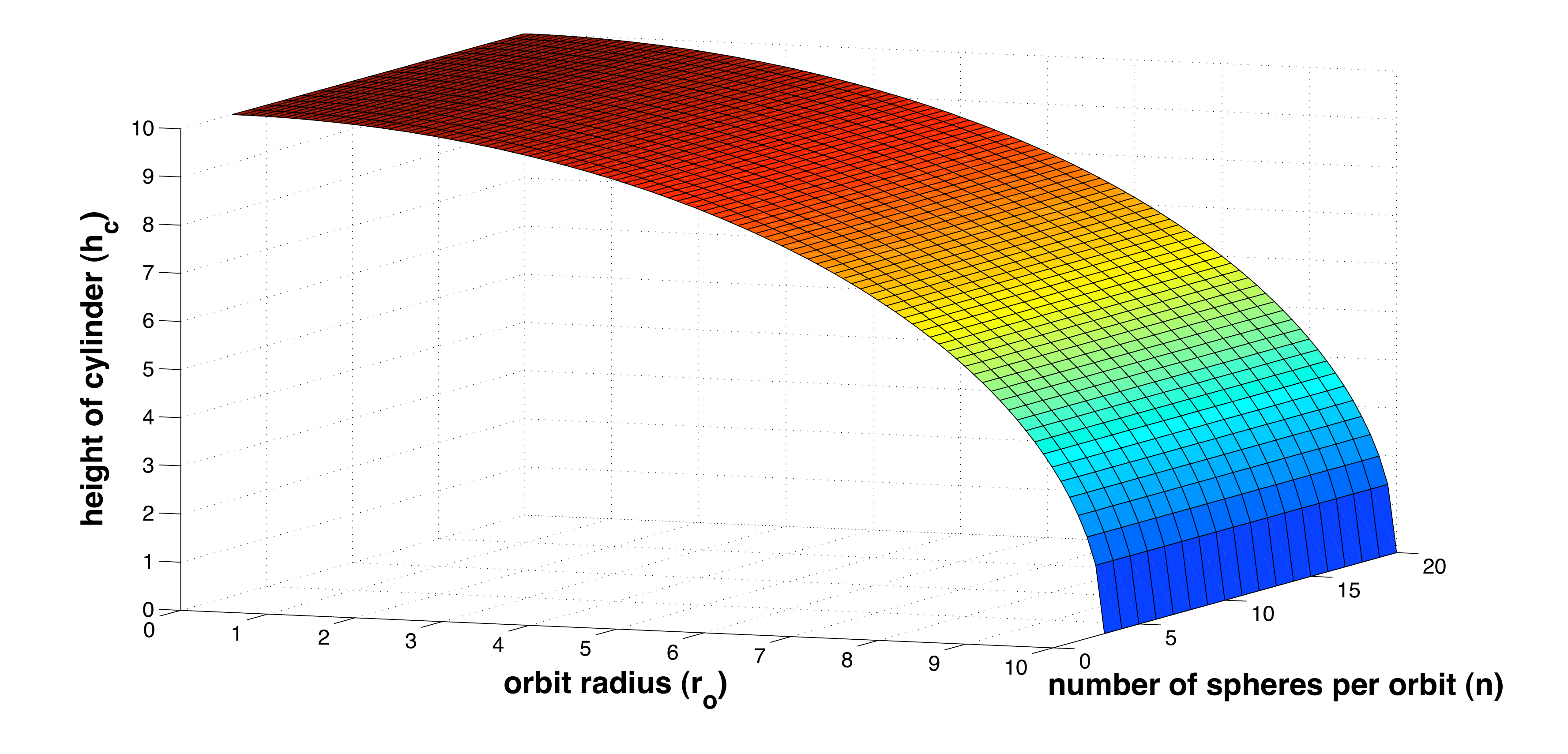}\label{fig:height}}
\hfill
\subfigure[Volume of Cylinder]{\includegraphics[width=0.325\textwidth,keepaspectratio]{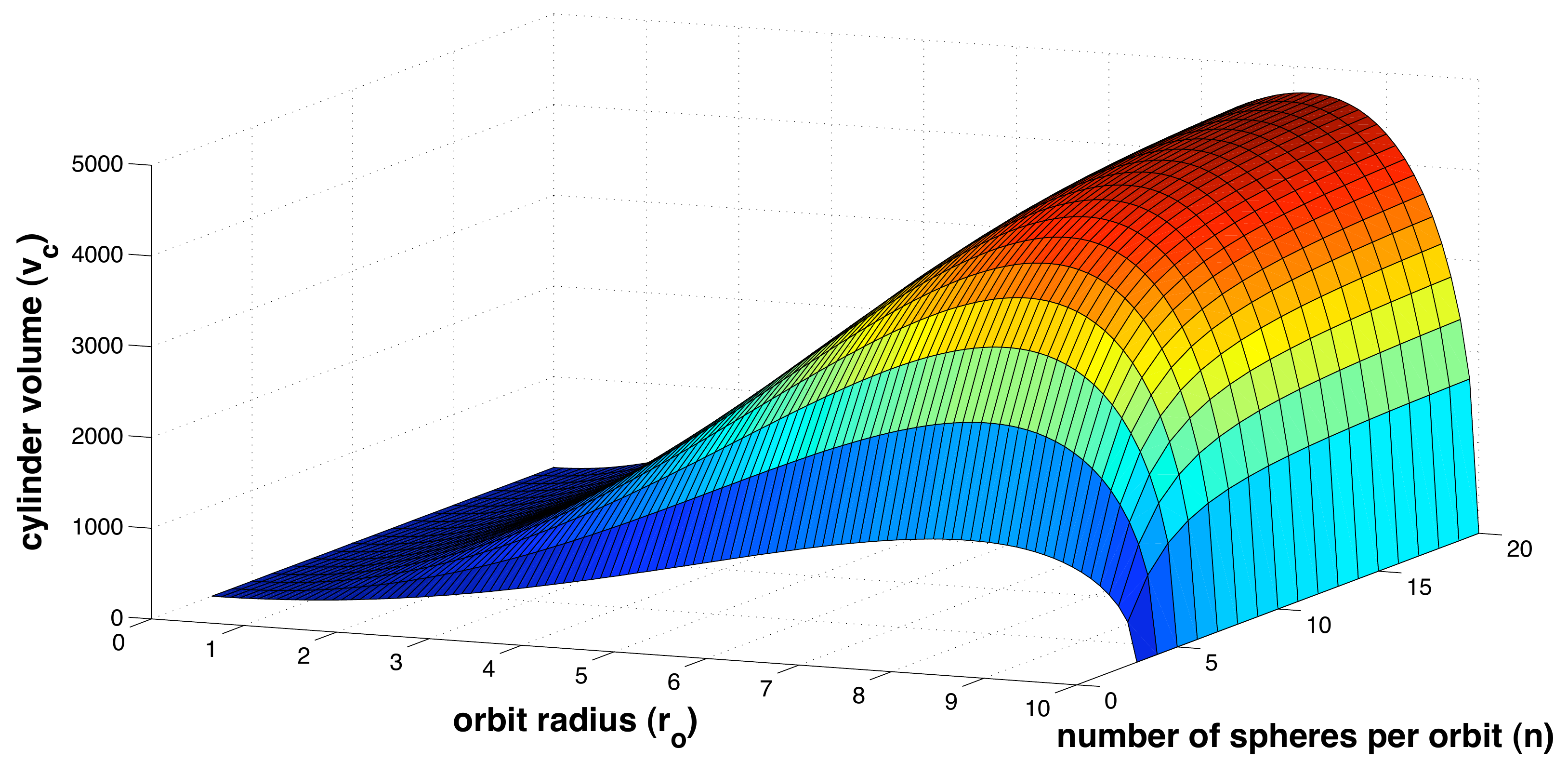}\label{fig:volume}}
\caption{Change of Cylinder Radius, Height and Volume with Orbit Radius and Number of Spheres per Orbit with fixed Sphere Radius = 10 units}
\label{fig:cylin}
\end{figure*}

Therefore, given the radius of the orbit $r_{o}$, the number of spheres moving in each orbit $n$, and radius of each sphere $r_{s}$, the cylinder can be determined by using the above equations as follows:
\begin{eqnarray}
\label{eq:rc}
r_{c} & = & 2(h_{l}-y) ~~~~~~~~~~~~~~~~~(\textnormal{from equation \ref{eq:rcy}}) \nonumber\\
& = & 2(r_{s}-w_{l}) cot \frac{\theta}{2} ~~~~~~~~~~(\textnormal{from equation \ref{eq:rswl}}) \nonumber\\
& = & 2 r_{o} sin \frac{\theta}{2} cot \frac{\theta}{2} ~~~~~~~~~~~~(\textnormal{from equation \ref{eq:rsro}}) \nonumber\\
r_{c} & =  & 2 r_{o} cos \frac{\theta}{2} = 2 r_{o} cos \frac{\pi}{n}
\end{eqnarray}
Using equations \ref{eq:hlwl}, \ref{eq:rswl}, \ref{eq:rsro} and \ref{eq:hcy}, we can express $h_{c}$ as :
\begin{eqnarray}
h_{c}^{2} & = & h_{l}^{2} - (h_{l}-y)^{2} \nonumber\\
& = & (r_{s}^{2} - (r_{s}-w_{l})^{2}) - ((r_{s}-w_{l})cot \frac{\theta}{2})^{2}\nonumber \\
& = & r_{s}^{2} - ((r_{s} - w_{l})\frac{1}{sin \frac{\theta}{2}})^{2} \nonumber\\
& = & r_{s}^{2} - r_{o}^{2} \nonumber\\
h_{c} & = & \sqrt{r_{s}^{2} - r_{o}^{2}}\label{eq:hc}
\end{eqnarray}
The volume of cylinder is given by:
\begin{eqnarray}
\label{eq:vc}
v_{c} & = & \pi r_{c}^{2} h_{c}
\end{eqnarray}

The change of the radius ($r_{c}$), height ($h_{c}$) and volume ($v_{c}$) of the cylindrical region with the values of $r_{o}$ and $n$ for fixed value of $r_{s}$ can be seen in the plots in Figs. \ref{fig:radius}, \ref{fig:height} and \ref{fig:volume}. The coverage problem for an airborne network can be formally defined as follows. Given:
\begin{itemize}
	\item The length, width and height of the air corridor as $L_{ac}$, $W_{ac}$ and $H_{ac}$.
	\item  The radius of each coverage sphere associated with each ANP $r_{s}$.
\end{itemize}
Find (i) the orbit radius, $r_{o}$, (ii) the number of ANPs in each orbit, $n$, (iii)  the number of orbits $m$ required to cover the air corridor, and (iv) the placement of the center of the orbits $(x_{i},y_{i}),~1\leq i\leq m$, such that:

\begin{enumerate}
	\item The orbits are placed only on the top surface of the air corridor.
	\item All points in the rectangular area defined by $L_{ac}$ and $W_{ac}$ is covered by at least one circle of radius $r_{c}$ with center at $(x_{i},y_{i}),~1\leq i\leq m$.
	\item $h_{c} \geq H_{ac}$
	\item Total number of spheres required = total number of orbits $\times$ number of spheres per orbit = $mn$ is minimized.
\end{enumerate}

\section{Design For Coverage - Solution}
\label{sec:coverageSolution}

\begin{figure}[!t]
\centering
\subfigure[Placement using Strategy 1]{\includegraphics[width=0.24\textwidth, keepaspectratio]{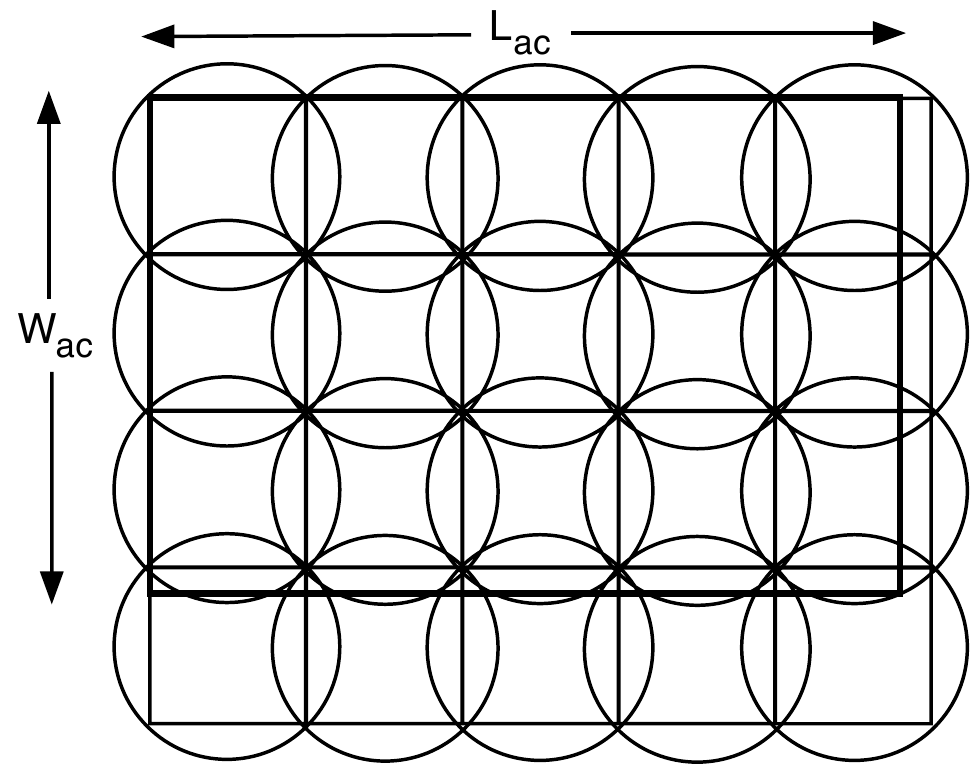}\label{fig:illStrat1}}
\hfill
\subfigure[Placement using Strategy 2]{\includegraphics[width=0.24\textwidth, keepaspectratio]{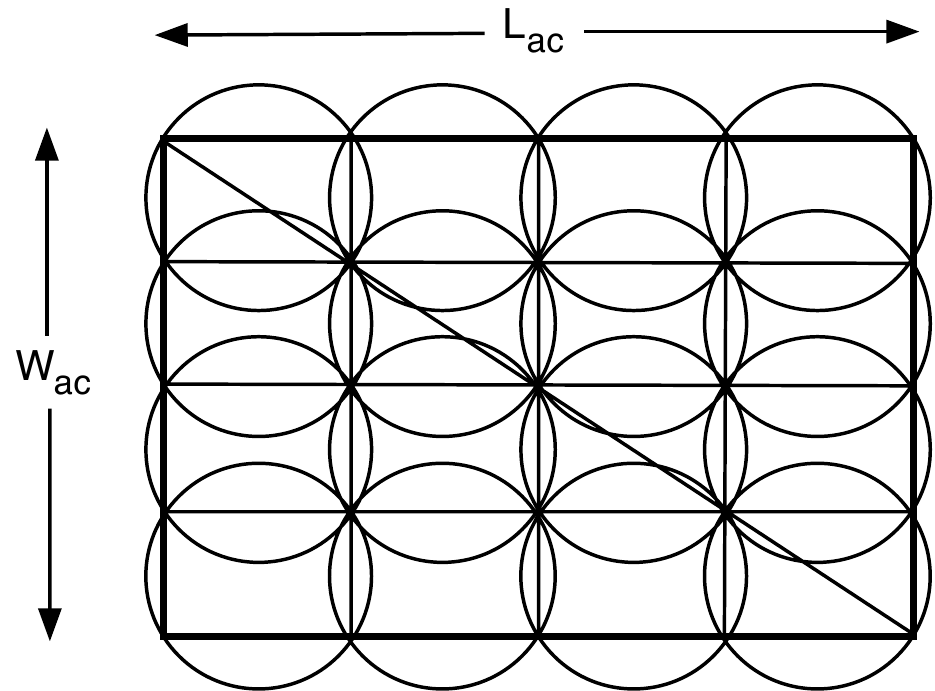}\label{fig:illStrat2}}
\caption{Covering $L_{ac}\times W_{ac}$ plane with circles of radius $r_{c}$ using strategy 1 and 2}
\label{fig:strats}
\end{figure}

In order to minimize the objective function subject to the constraints, we need to find $r_{o}$ and $n$. This will determine the values of $r_{c}$ and $h_{c}$. It can be seen from equations (\ref{eq:rc}) and (\ref{eq:hc}), that decreasing $r_{o}$, increases $h_{c}$, but decreases $r_{c}$. Intuitively, with smaller value of $r_{c}$, we will need more orbits ($m$) to cover the rectangular parallelopiped, which will eventually increase the number of ANPs required ($mn$). Therefore, for given $r_{s}$ we need to set $r_{o}$ at the highest possible value that still satisfies the constraint $h_{c} \geq H_{ac}$. For the corresponding $r_{c}$, $m$ will be determined by the placement strategy of the orbits on the top surface of the air corridor. The overall objective of the coverage problem is to minimize $mn$ subject to the constraint $h_{c} \geq H_{ac}$.  We use the following two strategies for placement of circular orbits on the top surface of the air corridor:

\noindent
{\bf Strategy 1:} The largest square that can be inscribed in the circular surface of the cylinder is used as the building block to cover the rectangular region defined by $L_{ac}$ and $W_{ac}$. With $r_{c}$ being the radius of the cylinder, the length of each side $a$ of the square is $\sqrt{2}r_{c}$. Therefore, total number of orbits can be calculated as $m=\lceil \frac{L_{ac}}{\sqrt{2}r_{c}}\rceil \times \lceil \frac{W_{ac}}{\sqrt{2}r_{c}}\rceil$. Hence, the optimization problem following strategy 1 can be formally stated as:
\begin{eqnarray}
minimize~mn  =  \lceil \frac{L_{ac}}{\sqrt{2}r_{c}}\rceil \times \lceil \frac{W_{ac}}{\sqrt{2}r_{c}}\rceil \times n &&\nonumber \\
= \lceil \frac{L_{ac}}{2\sqrt{2}r_{o}cos\frac{\pi}{n}}\rceil \times \lceil \frac{W_{ac}}{2\sqrt{2}r_{o}cos\frac{\pi}{n}}\rceil \times n &&\label{eq:s1obj} \\
subject~to:~~h_{c}  = \sqrt{r_{s}^{2} - r_{o}^{2}} \geq  H_{ac} && \label{eq:s1cons}
\end{eqnarray}
{\bf Strategy 2:} The largest rectangle that can be inscribed in the circular surface of the cylinder, and that has the same length to width ratio as that of the rectangular region defined by $L_{ac}$ and $W_{ac}$, is used as the building block for covering the entire region. Let $a$ and $b$ be the length and width of such a building block.  $r_{c}$ being the radius of the cylinder, we get the following:
\begin{eqnarray*}
\frac{a}{b} = \frac{L_{ac}}{W_{ac}} \textnormal{    and     }(2r_{c})^{2} = a^{2} + b^{2}
\end{eqnarray*}
From the above two relations we can get:
\begin{eqnarray*}
a = \frac{2r_{c}L_{ac}}{\sqrt{L_{ac}^{2} + W_{ac}^{2}}},~~b = \frac{2r_{c}W_{ac}}{\sqrt{L_{ac}^{2} + W_{ac}^{2}}}
\end{eqnarray*}
Since $r_{c} = 2r_{o}cos\frac{\pi}{n}$, total number of orbits (cylinders) required $m$ to cover the entire region can be calculated as:
\begin{eqnarray*}
m & = & \lceil \frac{L_{ac}}{a} \rceil \times \lceil \frac{W_{ac}}{b} \rceil \\
& = & \lceil \frac{\sqrt{L_{ac}^{2} + W_{ac}^{2}}}{4r_{o}cos\frac{\pi}{n}} \rceil \times \lceil \frac{\sqrt{L_{ac}^{2} + W_{ac}^{2}}}{4r_{o}cos\frac{\pi}{n}} \rceil\\
\end{eqnarray*}
Accordingly, the optimization problem following placement strategy 2 can be formally stated as:
\begin{eqnarray}
minimize~mn = \lceil \frac{\sqrt{L_{ac}^{2} + W_{ac}^{2}}}{4r_{o}cos\frac{\pi}{n}} \rceil^{2} \times n &&\label{eq:s2obj} \\
subject~to:~~h_{c}  = \sqrt{r_{s}^{2} - r_{o}^{2}} \geq  H_{ac} && \label{eq:s2cons}
\end{eqnarray}
The diagrams for the placement of orbits, and hence the cylindrical regions following the above two strategies, are shown in Fig.~\ref{fig:illStrat1} and \ref{fig:illStrat2}, respectively. Hence, the location of he center of the orbits $(x_{i},y_{i})$ can be easily determined.
It may be observed both placement strategy 1 and 2 formulates the coverage problem as a non-linear optimization problem. We used the non-linear constrained program solver Nimbus \cite{nimbus} to solve optimization problems following strategies 1 and 2. The results obtained from Nimbus is discussed in Section \ref{sec:experiments}.

\section{Design for Connected Coverage}
\label{sec:connCoverage}

In Section \ref{sec:coverageProbFormulation}, we discussed the three dimensional air corridor coverage problem with the ANPs. As the ANPs are mobile, the coverage volume associated one ANP is continuously changing with time. In Section \ref{sec:coverageSolution} we described techniques to find least cost solution to the air corridor coverage problem with mobile nodes. Earlier in Section \ref{sec:connectivity}, we discussed how to determine the velocity and subsequently the transmission range of the of the ANPs, so that resulting backbone network formed by the ANPs remain connected at all times. In this section, we discuss how to design a network of ANPs, so that (i) it remains connected at all times and (ii) it provides 100\% coverage to the air corridor at all times.

We provide a two phase solution to the connected coverage problem. In the first phase, using the techniques described in Section \ref{sec:coverageSolution}, we determine the number and orbit of the ANPs that will provide 100\% coverage to air corridor at all times. Once that is accomplished, in the second phase, using the techniques described in Section \ref{sec:connectivity}, we determine the the velocity and the transmission range of the ANPs so that  the backbone network formed by the ANPS remains connected at all times. 

\section{Visualization Tool for Airborne Network Design}
\label{sec:visualization}
\begin{figure}
\centering
\includegraphics[width=0.45\textwidth, keepaspectratio]{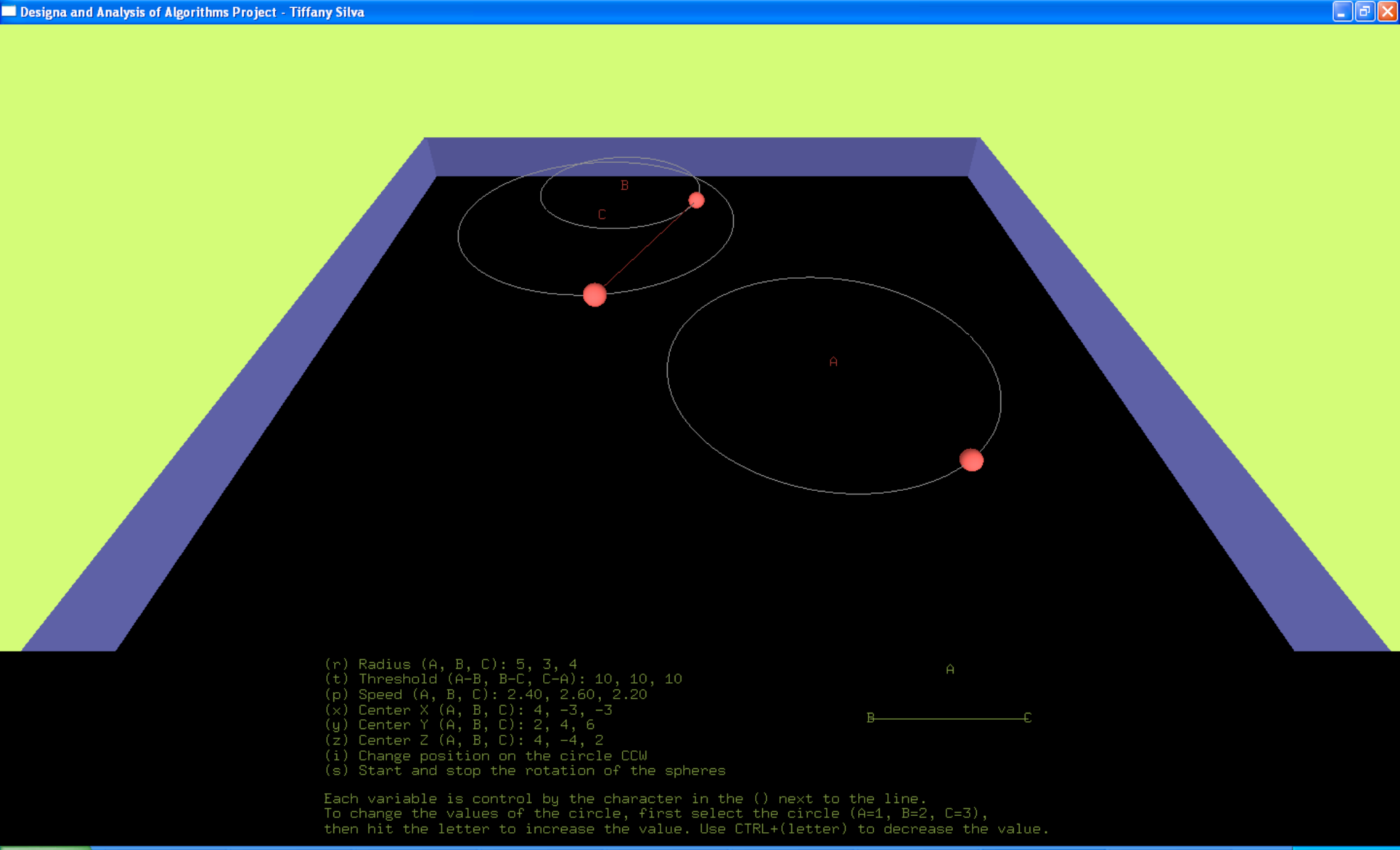}
\caption{Snapshot of Visualization Tool with Three Moving Objects}
\label{fig:viz}
\end{figure}

The visualization tool was designed for observing the movement of objects along circular orbits in a 3D plane. The Euclidean distance between every pair of objects keeps changing due the their movement. Each pair of objects has a threshold value specified. If the Euclidean distance between this pair of objectss is within the threshold, then they are connected by a link. As soon as the pairwise distance goes beyond the threshold, the link is broken. The tool was designed using OpenGL and C++. OpenGL is a 3D graphics API that works with C++ that provides dynamic interaction with the user. Mostly it is used for game programming and creating 3D scenes. For this program, we utilized some of the basic features of the API to create an interactive application to control the variables of each particular orbiting objects. One of the features of OpenGL is the ability to alter the camera view. This allows us to see every angle of the orbiting objects and rotate around the scene. A snapshot of the visualization tool with three moving points is shown in Fig.~\ref{fig:viz}.

The floor is based on a $24 \times 24$ grid. The center of each orbit has the ability to be moved along the $X$ axis, $-12 \leq X \leq 12$,  and the $Z$ axis $-12 \leq Z \leq 12$. The $Y$-axis, which represents the height above the floor, allows the object to increase in height in the range $1 \leq  Y \leq 10$. The radius of each circular orbit can be modified in the range $1 \leq R \leq 10$. Each object has a connectivity threshold to each other. Each threshold, $A-B$, $B-C$, and $C-A$ has the following range $1 \leq T \leq 20$. Using the standard distance equation between two points in $3D$ space, a link will appear to declare if the objects are within the given threshold limit. The speed of each object is based on the system clock, which will vary between each computer. The speed of each object can be increased up to $5$ units. Prior to each object being set in motion, they can be positioned strategically around their own orbit. The motion of all objects can be paused to analyze a specific situation. 

\section{Experimental Results and Discussion}
\label{sec:experiments}

\begin{figure*}
 \centering
 \subfigure[$L_{ac}=100,W_{ac}=70,H_{ac}=10$]{\includegraphics[width=0.32\textwidth, keepaspectratio]{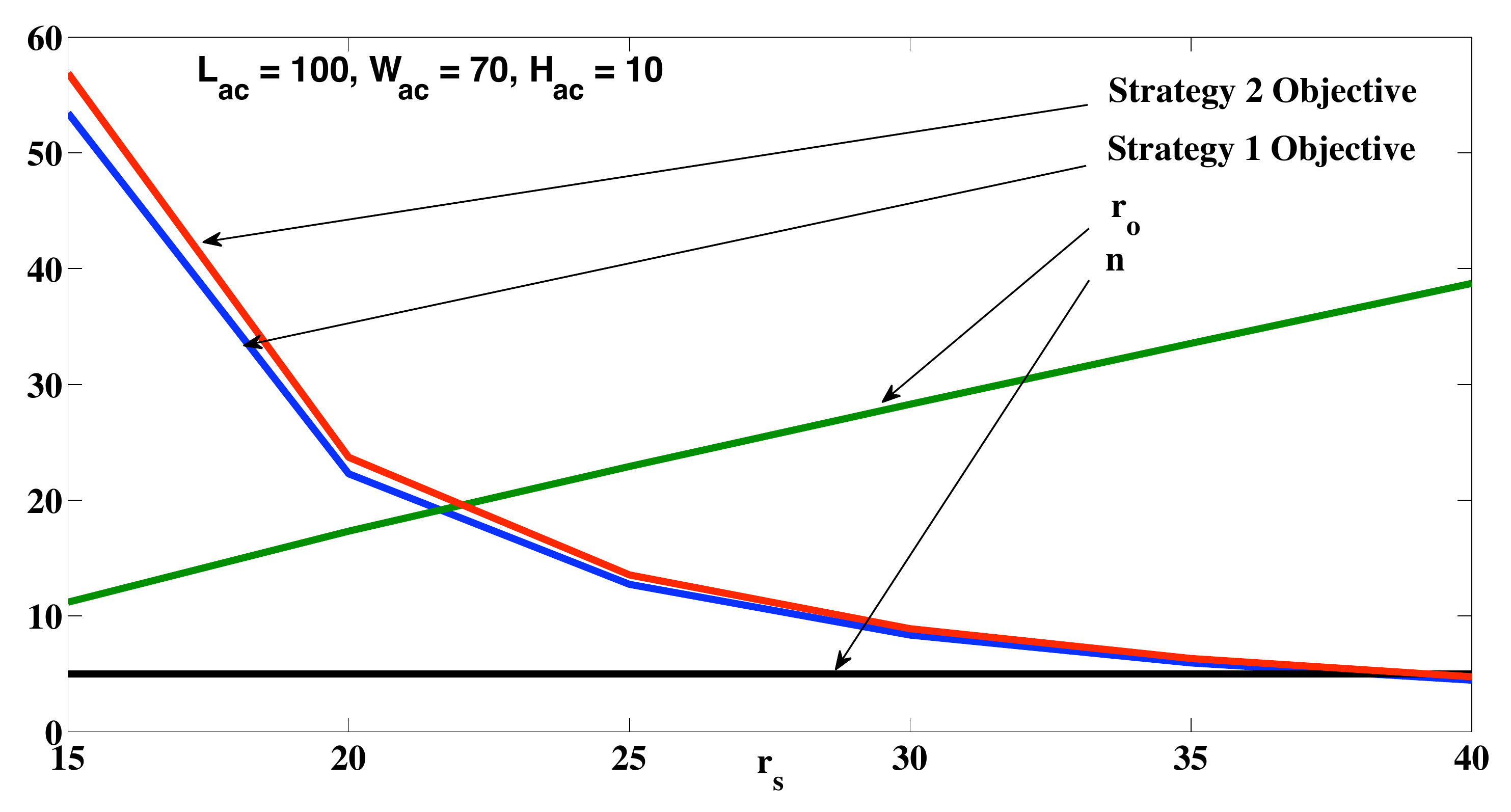}\label{fig:rs1}}
 \hfill
 \subfigure[$L_{ac}=100,W_{ac}=180,H_{ac}=10$]{\includegraphics[width=0.32\textwidth, keepaspectratio]{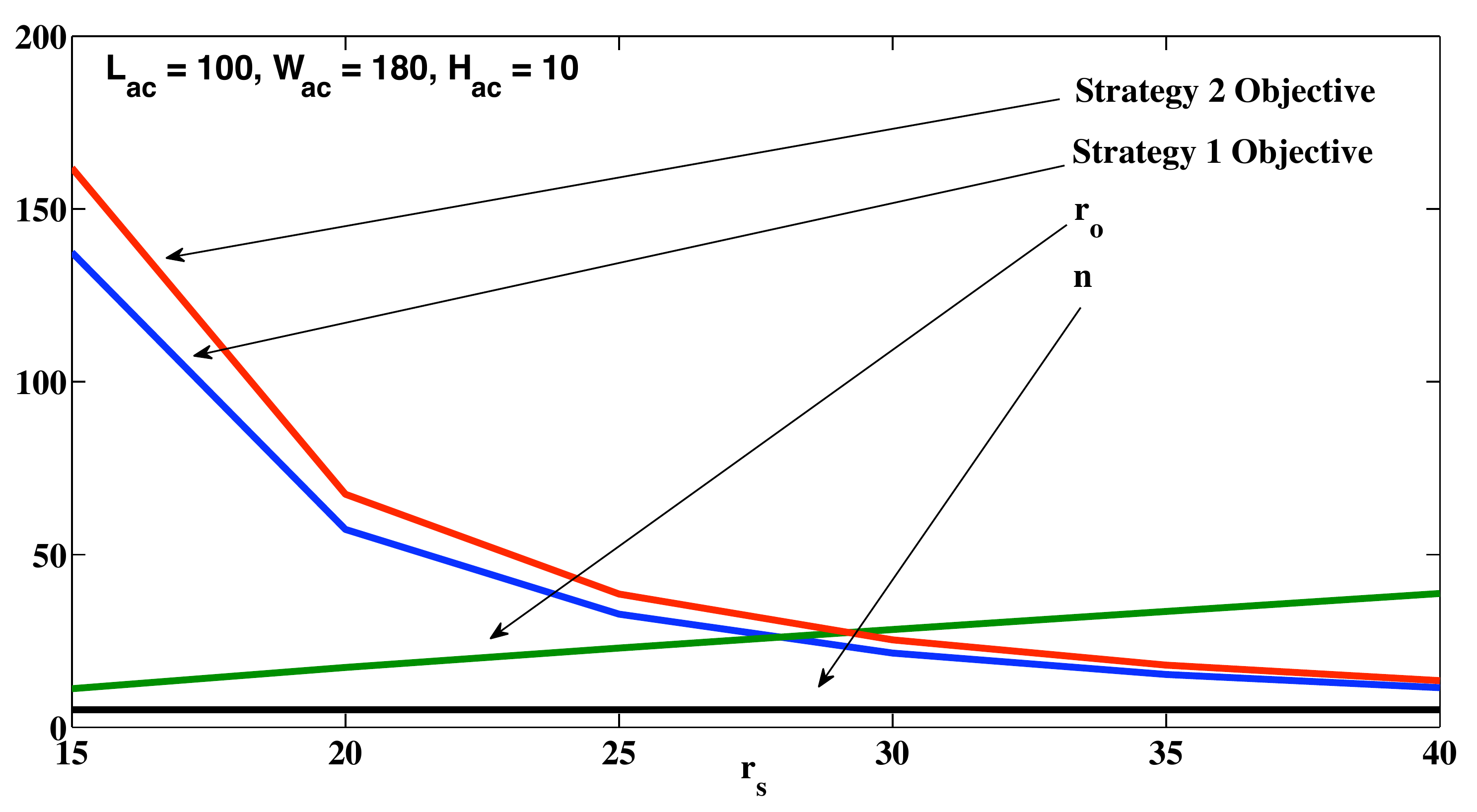}\label{fig:rs2}}
 \hfill
 \subfigure[$L_{ac}=100,W_{ac}=100,H_{ac}=10$: Strategy 1 becomes the same as Strategy 2]{\includegraphics[width=0.32\textwidth, keepaspectratio]{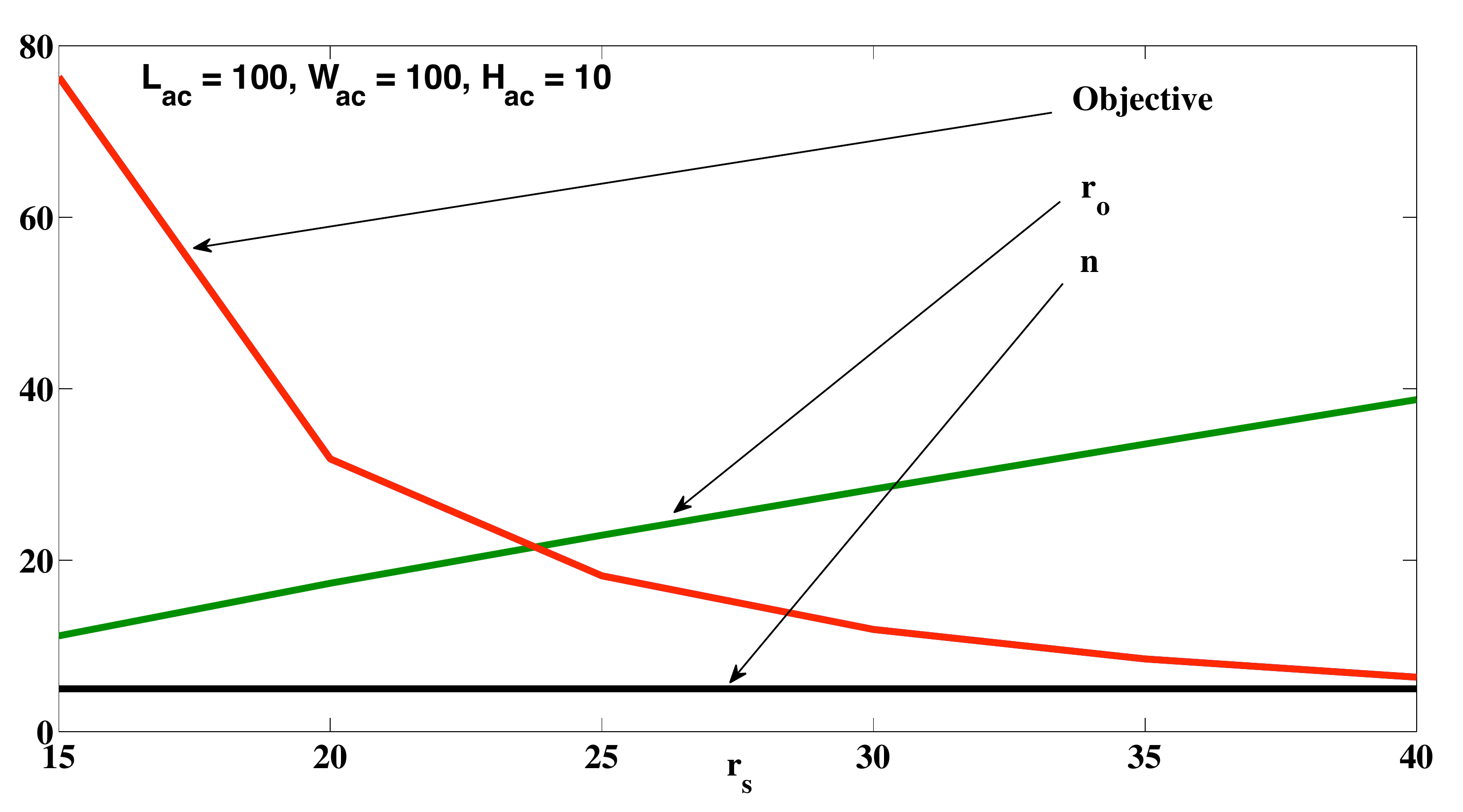}\label{fig:rs3}}
 \caption{Variation of the $r_{o}$, $n$ and the objective function for Strategy 1 and 2 (all values on $y$ axis) with variable $r_{s}$ and fixed value of $L_{ac},W_{ac}$ and $H_{ac}$}
 \label{fig:rs}
\end{figure*}

\begin{figure*}
 \centering
 \subfigure[$L_{ac}=100,W_{ac}=70,r_{s}=20$]{\includegraphics[width=0.32\textwidth, keepaspectratio]{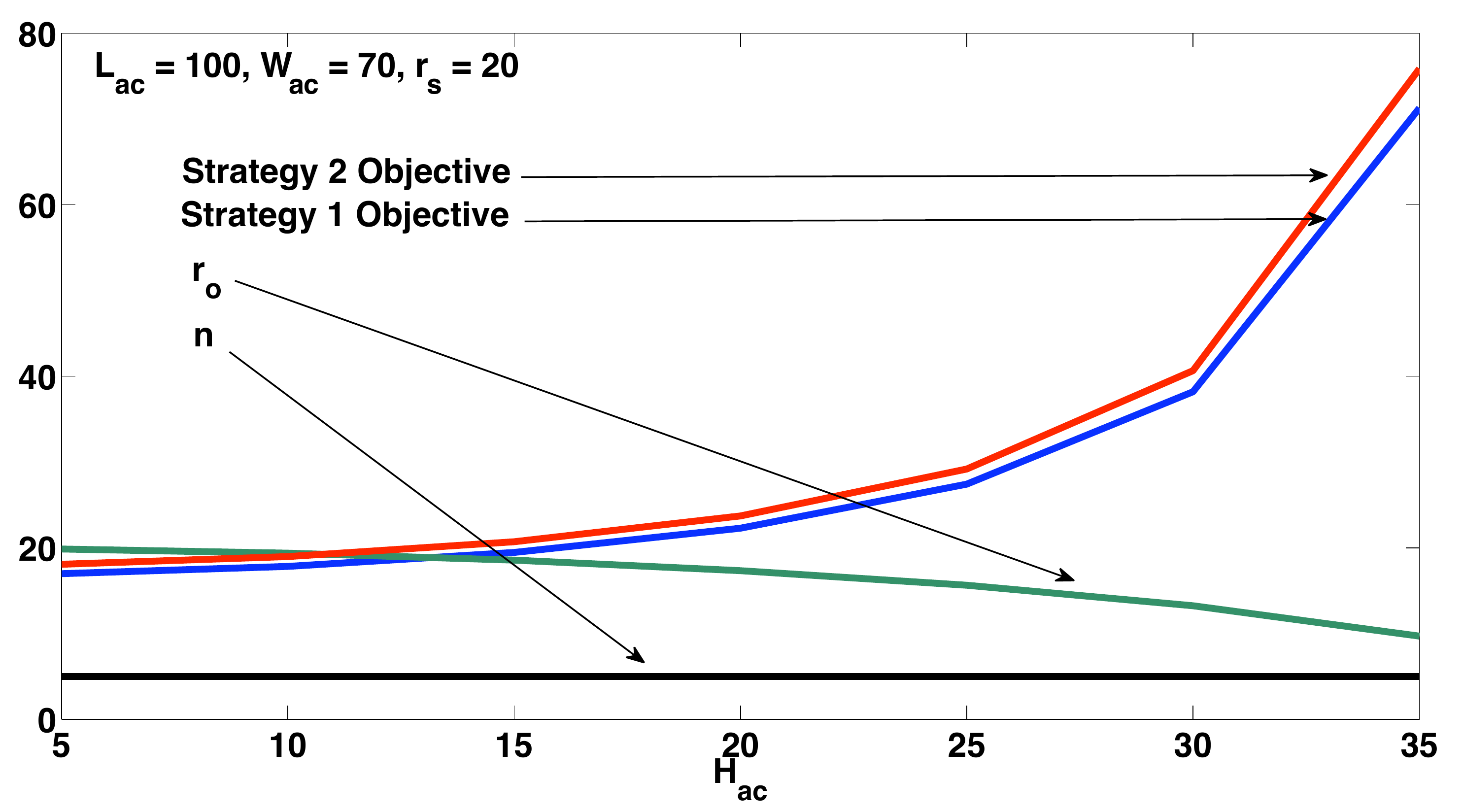}\label{fig:h1}}
 \hfill
 \subfigure[$L_{ac}=100,W_{ac}=180,r_{s}=20$]{\includegraphics[width=0.32\textwidth, keepaspectratio]{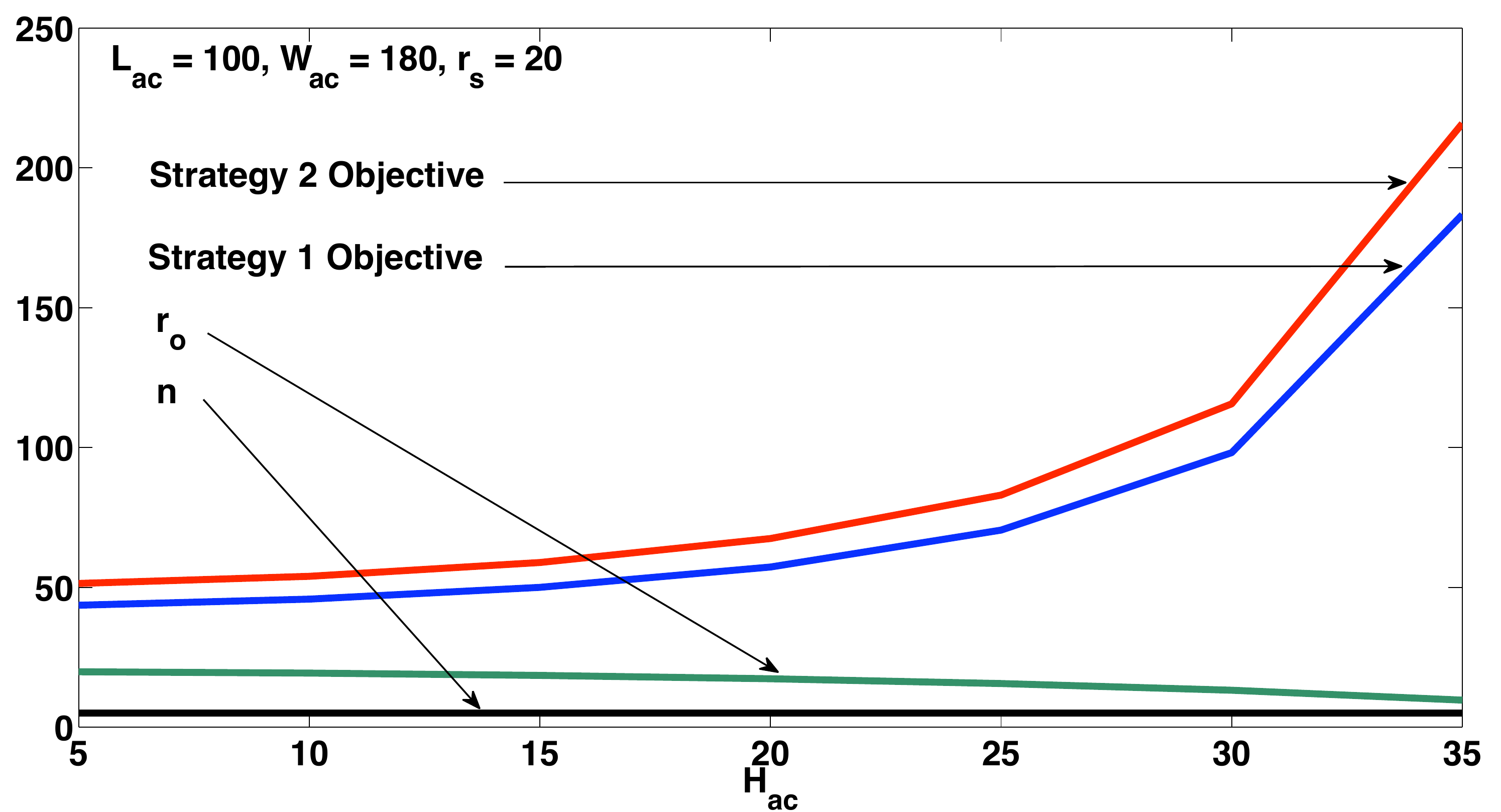}\label{fig:h2}}
 \hfill
 \subfigure[$L_{ac}=100,W_{ac}=100,r_{s}=20$: Strategy 1 becomes the same as Strategy 2]{\includegraphics[width=0.32\textwidth, keepaspectratio]{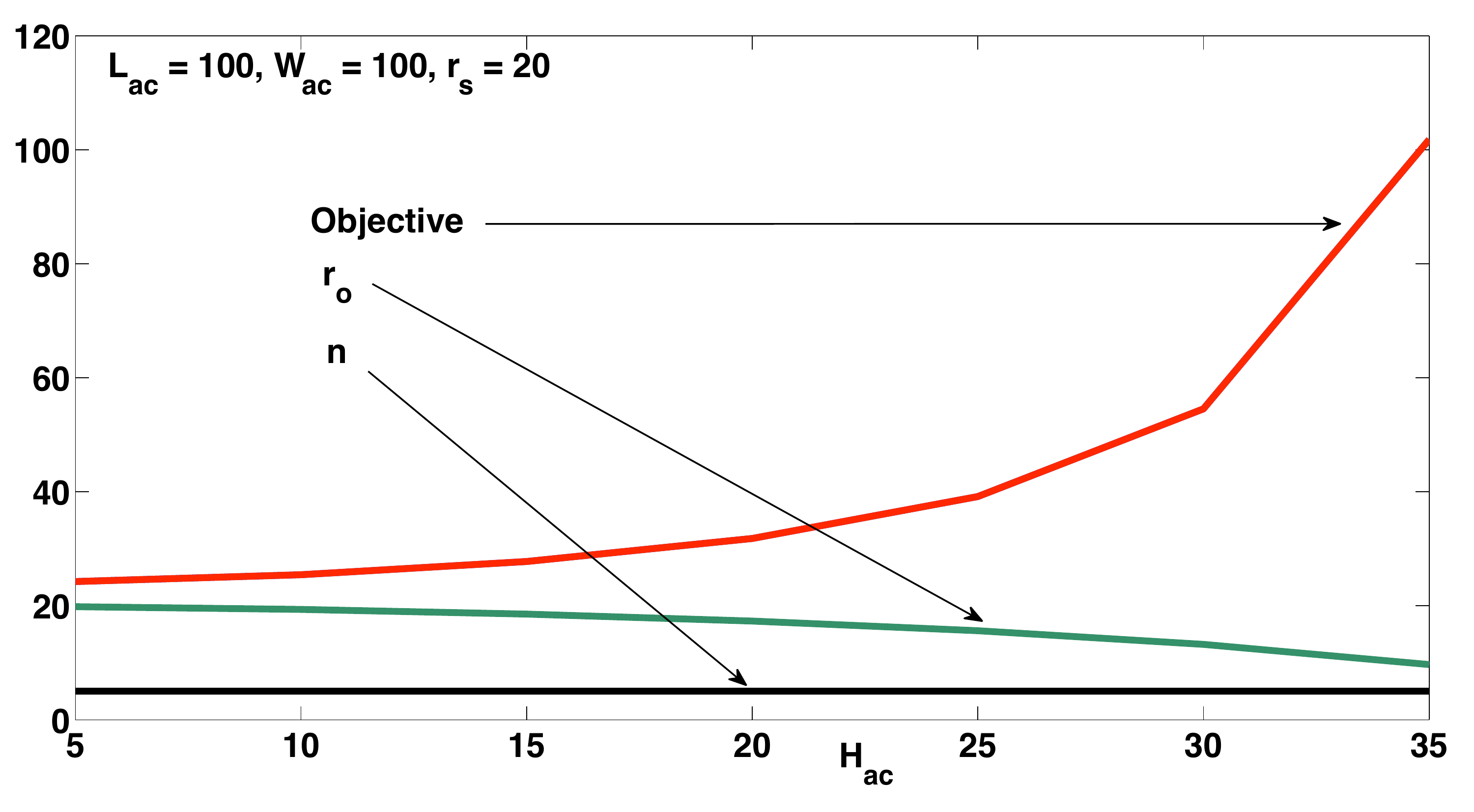}\label{fig:h3}}
 \caption{Variation of the $r_{o}$, $n$ and the objective function for Strategy 1 and 2 (all values on $y$ axis) with variable $H_{ac}$ and fixed value of $L_{ac},W_{ac}$ and $r_{s}$}
 \label{fig:h}
\end{figure*}

\begin{figure*}
 \centering
 \subfigure[$\frac{n}{cos^{2}(\frac{\pi}{n})}$ versus $n$]{\includegraphics[width=0.27\textwidth, keepaspectratio]{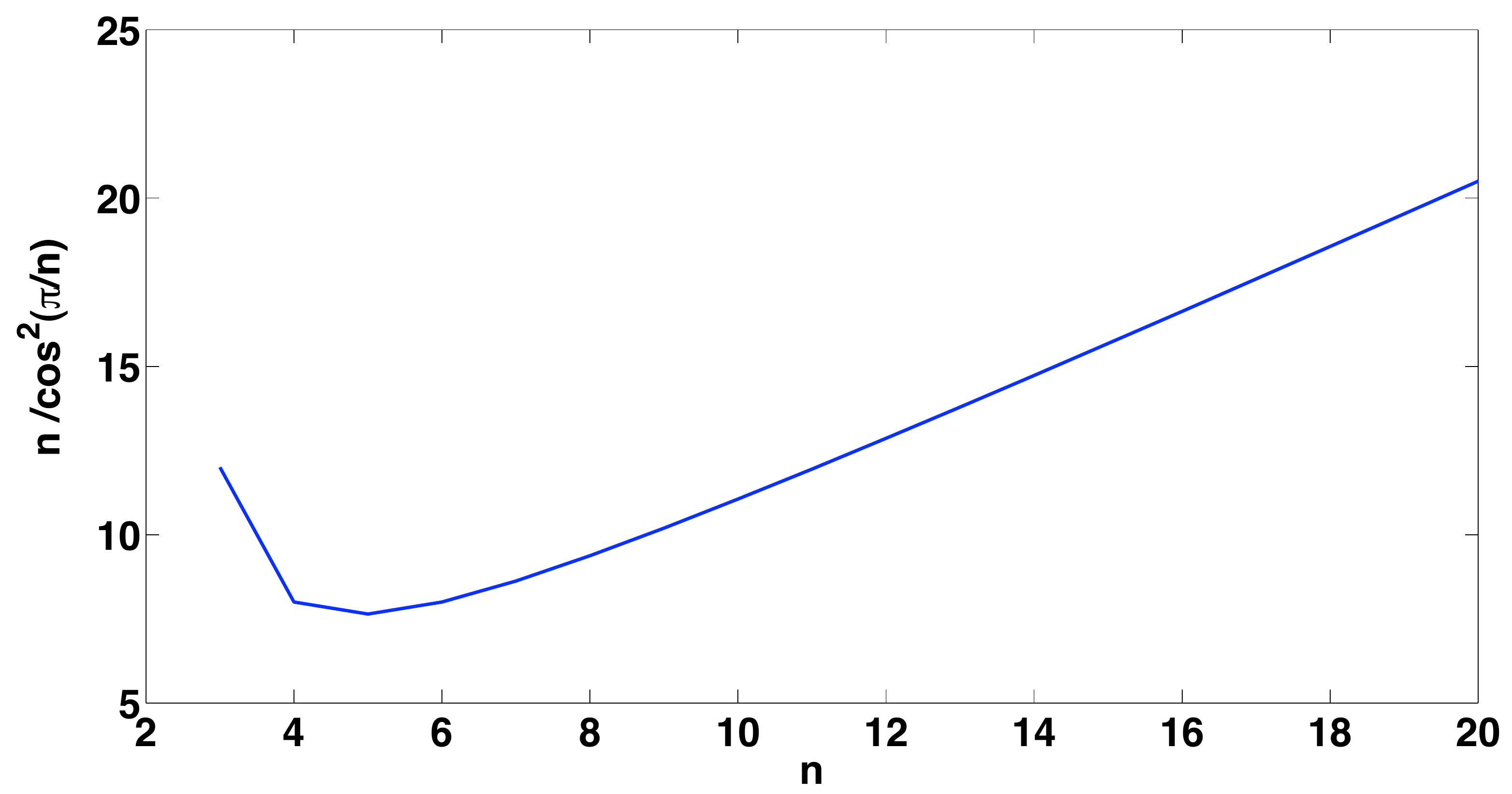}\label{fig:cos}}
 \hfill
 \subfigure[Change of objective function with $n$ and $r_{o}$]{\includegraphics[width=0.33\textwidth, keepaspectratio]{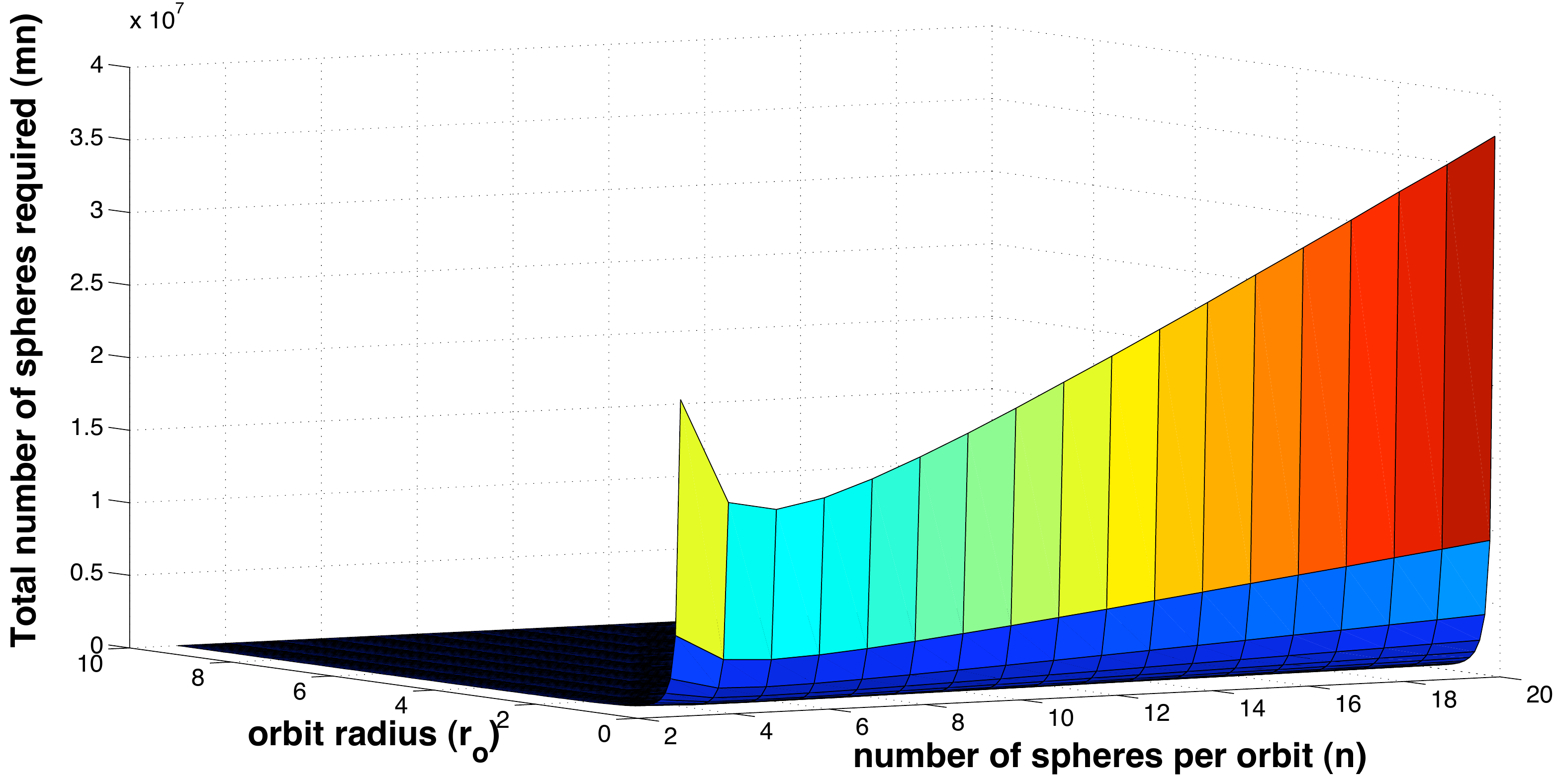}\label{fig:obj}}
 \hfill
 \subfigure[Objective function in log scale]{\includegraphics[width=0.35\textwidth, keepaspectratio]{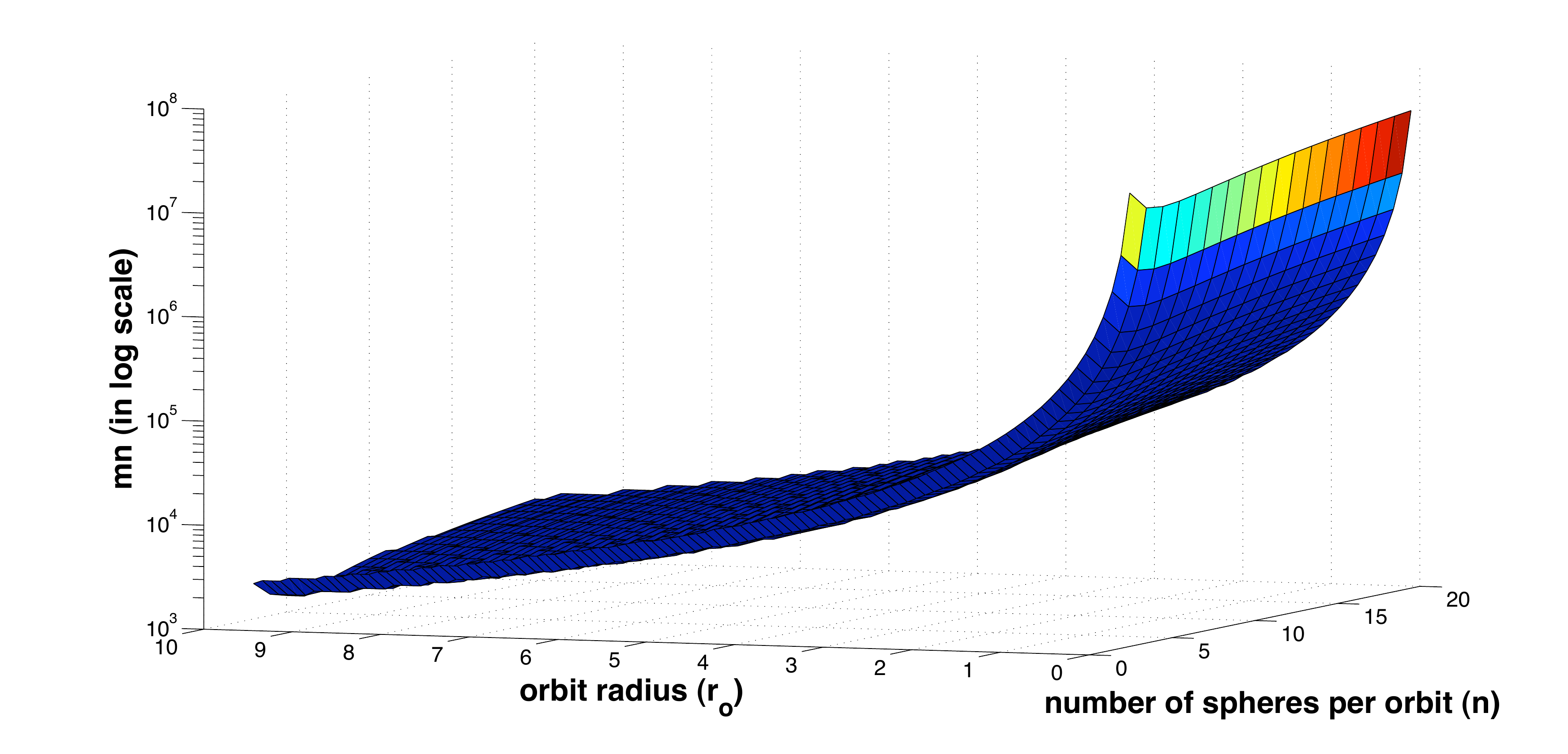}\label{fig:objLog}}
 \caption{The change of objective function and its components with $n$ and $r_{o}$}
 \label{fig:objs}
\end{figure*}

In this section we present the experimental evaluation results of two strategies proposed in Section \ref{sec:coverageSolution}.  The goal of these experiments were to find the impact of change of (i) radius of the coverage sphere ($r_s$) and (ii) height of the air corridor ($H_{ac}$) on (a) radius  of the circular orbit of the flying ANPs ($r_o$), (b) the number of ANPs in each orbit ($n$), and (c) the total number of ANPs ($mn$) needed to provide complete coverage for the air corridor, specified by its length, width and height parameters $L_{ac}, W_{ac}, H_{ac}$, respectively.  Fig.~\ref{fig:rs} show impact of changing $r_s$ on $r_o, n$ and $mn$ for different sets of values for $L_{ac}, W_{ac}, H_{ac}$ and for two different strategies 1 and 2. Fig.~\ref{fig:h} show impact of changing $H_{ac}$ on $r_o$, $n$ and $mn$ for different set of values for $L_{ac}, W_{ac}$, $r_s$ and for two different strategies 1 and 2. The parameter values for $L_{ac}, W_{ac}, H_{ac}$ and $r_s$ used for the experimentation are indicated in the figures.

Since the optimal coverage problem turned out to be a non-linear optimization problem (equations (\ref{eq:s1obj}) to (\ref{eq:s2cons})), we used the non-linear constrained program solver Nimbus \cite{nimbus} to solve it using two different ANP orbit placement strategies 1 and 2. In the following we discuss some experimental results, some of which are intuitive, some others are not.

\medskip
\noindent
{\em Observation 1: }From Fig.~\ref{fig:rs}, it can be seen that increase in $r_s$ results in increase in $r_o$ and decrease in $mn$ for both the strategies 1. This is somewhat intuitive as it is only natural to expect that as the radius of the coverage sphere increases, the radius of the circular orbit of the ANPs will increase and the total number of ANPs needed to cover the entire air corridor will decrease. It may also be noted that when $r_s$ is too small compared to $H_{ac}$, there may not be a feasible solution.
    
\medskip
\noindent
{\em Observation 2: }From Fig.~\ref{fig:h}, it can be seen that increase in $H_{ac}$ results in decrease in $r_o$ and increase in $mn$ for both the strategies 1 and 2. This is also somewhat intuitive,  as the height of the air corridor increases, the radius of the circular orbit of the ANPs has to decrease (please see discussion in Section \ref{sec:coverageSolution}) and the total number of ANPs needed to cover the entire air corridor must increase. 

\medskip
\noindent
{\em Observation 3: }From Figs.~\ref{fig:rs} and \ref{fig:h}, it can be seen that, $n$, the number of ANPs in an orbit remains a constant irrespective of changes in $L_{ac},W_{ac}, H_{ac}$ and $r_{s}$. This result is not at all obvious. However, on closer examination of the objective function in equations (\ref{eq:s1obj}) and (\ref{eq:s2obj}), one can find an explanation for this phenomenon. The plot of $n/cos^{2}(\frac{\pi}{n})$ versus $n$ is shown in Fig.~\ref{fig:cos}. This factor is present in the objective function for both the strategies. From Fig.~\ref{fig:cos}, $n/cos^{2}(\frac{\pi}{n})$ reaches its minimum value when  $n=5$. Therefore the objective functions in equations (\ref{eq:s1obj}) and (\ref{eq:s2obj}) are minimized when $n = 5$. This nature of $n$ also explains the fact in observation 1, where $mn$ decreases when $m$ decreases (i.e., when $r_o$ increases). Similarly, it also explains the fact in observation 2, where $mn$ increases when $m$ increases (i.e., when $r_o$ decreases).
 
\medskip          
\noindent
 {\em Observation 4: }From the Figs.~\ref{fig:rs1},~\ref{fig:rs2},~\ref{fig:h1},~\ref{fig:h2}, it can be seen that the cost of the solution (i.e., the number of ANPs needed to provide complete coverage of the air corridor) using strategy 1 is less than that of strategy 2. Although, the reason for this phenomenon may not be obvious at a first glance, on closer examination, we can explain the phenomenon.  Given the fact that $L_{ac}^{2} + W_{ac}^{2} \geq 2L_{ac}W_{ac}$ and presence of these two terms in objective functions of strategies 1 and 2 (equations (\ref{eq:s1obj}) and (\ref{eq:s2obj}) in page 8), it is not surprising that cost of the solution strategy 1 is less than that of strategy 2.

From our experiments we learn that (i)strategy 1 performs better than strategy 2 in all cases, except where $L_{ac} = W_{ac}$, for which both the strategies are identical, (ii)the number of ANPs in an orbit remains a constant ($5$) irrespective of the values of $L_{ac},W_{ac}, H_{ac}$ and $r_{s}$, when the objective function is specified  by equations (\ref{eq:s1obj}) or (\ref{eq:s2obj}), and (iii)to optimize the objective function, the radius of the circular orbit of the ANPs should be made as large as possible subject to the constraint that the height of the corresponding invariant coverage cylinder is at least as large as the height of the air corridor $H_{ac}$.

\section{Conclusion}
\label{sec:conclusion}
\balance Existence of sufficient control over the movement pattern of the mobile platforms in Airborne Networks opens the avenue for designing topologically stable hybrid networks. In this paper, we discussed the system model and architecture for Airborne Networks (AN). We studied the problem of maintaining the connectivity in the underlying dynamic graphs of airborne networks with control over the mobility parameters and developed an algorithm to solve the problem. 

\bibliographystyle{IEEEtran}


\end{document}